\documentclass[11pt]{article}
\usepackage{mydef2col}
\usepackage{myArticle}

\title{American options valuation in time-dependent jump-diffusion models via integral equations and characteristic functions}
\author{
    \authorstyle{ Andrey Itkin}
    \newline \newline
    \institution{FREE department, Tendon School of Engineering, New York University, email: \url{aitkin@nyu.edu}} \\
}

\date{}

\begin{document}

\maketitle.

\vspace{-1.in}
\lettrineabstract{Despite significant advancements in machine learning for derivative pricing, the efficient and accurate valuation of American options remains a persistent challenge due to complex exercise boundaries, near-expiry behavior, and intricate contractual features. This paper extends a semi-analytical approach for pricing American options in time-inhomogeneous models, including pure diffusions, jump-diffusions, and \LY processes. Building on prior work, we derive and solve Volterra integral equations of the second kind to determine the exercise boundary explicitly, offering a computationally superior alternative to traditional finite-difference and Monte Carlo methods. We address key open problems: (1) extending the decomposition method, i.e. splitting the American option price into its European counterpart and an early exercise premium, to general jump-diffusion and \LY models; (2) handling cases where closed-form transition densities are unavailable by leveraging characteristic functions via, e.g., the COS method; and (3) generalizing the framework to multidimensional diffusions. Numerical examples demonstrate the method's efficiency and robustness. Our results underscore the advantages of the integral equation approach for large-scale industrial applications, while resolving some limitations of existing techniques.
}


\vspace{1em}

Despite the remarkable recent success of machine learning in pricing and calibrating various derivative products, training neural networks (even offline) requires fast and accurate methods for option pricing. For American options, this presents a persistent challenge due to several factors: the complex structure of exercise boundary(ies), the special behavior of options near expiry, and the inherent nature of option contracts, particularly for assets with discrete dividend payments, complex coupon schedules, and other contractual features. Consequently, developing efficient numerical methods, and more importantly, semi-analytical pricing algorithms, remains a critical and active area of research.

This paper continues a series of the author's works on the semi-analytical pricing of American options in models where the underlying asset follows time-inhomogeneous dynamics, which may be a pure diffusion, jump-diffusion, or a \LY process.

The foundation of this approach was first introduced in \cite{CarrItkin2020jd}. Subsequently, in \cite{ItkinMuravey2024jd}, the methodology was extended to various popular one-factor time-inhomogeneous diffusion models, incorporating the computation of the exercise boundary (EB) by solving a specific class of Volterra integral equations of the second kind. The paper \cite{ItkinMuravey2024jd} derived these integral equations for each model under consideration and explored numerical methods for their efficient solution. Subsequently, in \cite{Itkin2024jd}, this methodology was extended to jump-diffusion models featuring single and double exponential jumps.

In \cite{ItkinKitapbayev2025}, an alternative Volterra integral equation of the second kind was derived by extending the earlier framework of \cite{Peskir2005}. For pure diffusion models, this approach decomposes the American option price into its European counterpart and an early exercise premium (EEP). The EEP is expressed as a temporal integral of the risk-neutral expectation of a time-inhomogeneous function (explicitly determined by the underlying asset's stochastic dynamics) over the exercise region. The unknown exercise boundary (EB) can then be determined either by substituting the known option value at the EB into this equation and solving the resulting nonlinear algebraic equation, or by applying the methodology developed in \cite{ItkinMuravey2024jd}.

To emphasize, semi-analytical methods for pricing American options offer a compelling alternative to traditional numerical techniques, such as finite-difference (FD) methods or least-squares Monte Carlo (LSMC). Unlike conventional approaches, which implicitly define the EB while simultaneously solving for the option price, this framework allows for explicit determination of the EB's location. A detailed discussion of the challenges associated with traditional American option pricing methods can be found in \cite{Andersen2016}. Notably, the authors of \cite{Andersen2016} advocate for the integral equation approach in the Black-Scholes model with constant coefficients, demonstrating that it admits an efficient numerical scheme with convergence rates several orders of magnitude faster than standard FD and tree-based methods.

As highlighted in \cite{Itkin2024jd}, this methodology provides substantial computational advantages for industrial applications involving large-scale American option pricing. This observation aligns with \cite{Andersen2016} findings, which reveal persistent challenges in FD and LSMC methods, challenges that typically require increasingly sophisticated algorithmic solutions rather than uniform fixes. The authors particularly stress that creating robust, high-performance numerical methods for American options continues to be an important research frontier, further validating the superior efficiency and accuracy of the integral equation approach for the EB determination compared to traditional PDE-based (partial differential equation) numerical solutions.

Despite the advantages of the integral equation approach, several open problems remain that warrant further investigation:

\begin{itemize}
\item While the method in \cite{ItkinKitapbayev2025} was developed for pure diffusion models, its applicability to general time-dependent jump-diffusion models and \LY processes remains unexplored.

\item Calculating risk-neutral expectations for the EEP requires knowledge of the transition density. However, for many time-inhomogeneous models, such densities are unavailable in closed form. As shown in \cite{ItkinKitapbayev2025}, an alternative approach based on \cite{ItkinMuravey2024jd} may circumvent this issue.

\item Many widely used financial models lack closed-form transition densities but possess known characteristic functions. Extending the current methodology to leverage characteristic functions would significantly broaden its applicability.
\end{itemize}

In this paper, we tackle these challenges and show that a positive resolution is achievable in all the aforementioned cases.

The rest of the paper is organized as follows: \cref{S:decomp} briefly presents the decomposition of the American option price into the European price and the EEP. It also discusses two methods for computing the EB: the first involves solving a Volterra integral equation of the second kind derived directly from the decomposition formula, while the second employs a similar equation (or a system of equations) obtained using the Generalized Integral Transform (GIT) method from \cite{CarrItkin2020jd,ItkinMuravey2024jd}. \cref{s:CF} explains how the transition density required for computing expectations in the decomposition formula can be represented via the characteristic function (CF) using the COS method of \cite{FangOosterlee2008}. \cref{s:nDim} generalizes the decomposition formula for multidimensional diffusion processes and reviews prior work on the subject. \cref{s:Jump}  extends the analysis to models incorporating jumps. \cref{s:numEx} provides several numerical examples demonstrating the application of the proposed method for pricing American options. The final section summarizes the findings and implications of the study.

\section{Pure diffusion models with a known transition density} \label{S:decomp}

In this section we follow the description of the problem provided in \cite{ItkinKitapbayev2025}. It is known, that the price $P(t,x)$ of, e.g.,  an American Put option on the underlying asset $X_s$ at time $s > t \geq 0$, with strike price $K$ and maturity $T \geq s$, conditional on $X_t = x$, solves the following optimal stopping problem \cite{CarrJarrowMyneni1992}:
\begin{equation}\label{prob-interest}
P(t,x) = \sup_{\tau \in \mathcal{T}_{t,T}} \mathbb{E}^{\mathbb{Q}} \left[ D(t,\tau)(K - X_\tau)^+ \,\Big|\, X_t = x \right],
\end{equation}
\noindent where $D(t,s) = e^{-\int_t^s r(u)\,du}$ is the deterministic discount factor, $r(t)$ denotes the instantaneous risk-free interest rate, and $\mathbb{E}^{\mathbb{Q}}$ represents the expectation under the risk-neutral measure $\mathbb{Q}$. The supremum is taken over the set $\mathcal{T}_{t,T}$ of all $\mathcal{F}^X$-stopping times $\tau$ in the interval $[t,T]$, where $\mathcal{F}^X$ is the natural filtration generated by the underlying process $X$.

For $\forall x \in l_x$ we distinguish the exercise ($\mathcal{E}$) and continuation ($\mathcal{C}$) regions as
\begin{align}
\mathcal{E} &= \Big\{ (u,X_u)\in[0,T)\times (l_x,\infty): \, V(u,X_u) = K - X_u \Big\} \\
\mathcal{C} &= \Big\{ (u,X_u)\in[0,T)\times (l_x,\infty): \, V(u,X_u) > K - X_u \Big\}, \nonumber
\end{align}
\noindent where $l_x$ is the left boundary of the $X_t$ domain which could be either $l_x = 0$ or $l_x = -\infty$ depending on the model. These two regions are separated by the early EB $X_B(t)$, which is a time-dependent function of the time $t$. Note, that in this paper we don't consider the case of multiple EBs as the latter is discussed in detail in \cite{ItkinKitapbayev2025}.

We proceed with the decomposition of the American option price into several components, first proposed in \cite{CarrJarrowMyneni1992}, where the value of an American Put option is represented as a sum of the corresponding European Put price and the EEP. While the European option component maintains a consistent form across all cases, the early exercise premium depends on both the optimal exercise boundary (or boundaries) and the structure of the exercise region.

In this section we consider pure diffusion models of the type
\begin{equation} \label{model}
dX_t = \mu(t, X_t) dt + \Sigma(t, X_t) dW_t,
\end{equation}
where $W_t$ is the Brownian motion under the risk-neutral measure $\mathbb{Q}$, $\mu(t, X_t)$ is the drift, $\Sigma(t, X_t)$ is the diffusion coefficient, both the drift and volatility are some known deterministic functions of $(t,X_t)$. Then generalization of the decomposition of \cite{CarrJarrowMyneni1992} was provided in \cite{ItkinKitapbayev2025} which in the case of a single EB is given by the following Proposition
\begin{proposition} \label{prop1}
Conditional on $X_t = x$, the American Put price with a {\bf single} exercise boundary $X_B(t)$ can be represented by the following decomposition formula
\begin{align} \label{decompGen}
P \left(t, x \right) &= \EQ \left\{ D(t,T) [K - X_T]^+ | X_T = x\right\} + \int_t^T D(t,u) \EQ \left\{ \left[ r(u)(K-X_u) + \mu(u,X_u) \right] \mathbf{1}_{X_u \in \mathcal{E}} \right\} d u.
\end{align}
Here, the first term represents the usual European Put price $P_E\left(t, x \right)$ while the second term is the EEP which depends on the early exercise boundary $X_B(t)$.
\begin{proof}[{\bf Proof}]
See Proposition~1 in \cite{ItkinKitapbayev2025}.
\end{proof}
\end{proposition}

Using $\psi\left(X_u, u | x, t \right)$ to denote the transition density function of $X_u$ conditional on $X_t = x$, we can rewrite the Put pricing formula \eqref{decompGen} as
\begin{gather} \label{decompDensity}
\resizebox{0.95\textwidth}{!}{$
P \left(t, x \right) = D(t,T) \int_0^K \left(K - X_T\right) \psi\left(X_T, T | x, t \right) d X_T + \int_t^T D(t,u) \int_0^{X_B(u)} H(u, X_u) \psi \left(X_u, u | x, t \right)  d X_u d u,
$}
\end{gather}
\noindent where $H(u, X_u) = r(u)(K-X_u) + \mu(u,X_u)$. For a given model of the underlying asset, the American Put option price can be explicitly represented as in \eqref{decompDensity} and computed when both the transition density and the EB are known. When the second integral in \eqref{decompDensity} (the EEP) is positive,
the American Put value exceeds its European counterpart. Otherwise, early exercise is never optimal, and the American and European option prices coincide.

The formulation in \eqref{decompDensity} naturally accommodates models with time-dependent coefficients. For instance, when the drift takes the form $\mu(t, X_t) = [r(t) - q(t)] X_t$, where $q(t)$ is the dividend yield, the function $H(u, X_u)$ becomes
\begin{equation} \label{Htd}
H(u, X_u) = r(u) K - q(u) X_u,
\end{equation}
\noindent while \eqref{decompDensity} maintains its structure. This extension has important implications: since all model parameters become time-dependent, $H(u, X_u)$ and the EEP may change sign multiple times throughout the option's life. This behavior necessitates careful analysis of the exercise boundary topology when evaluating \eqref{decompDensity}; for detailed treatment, see \cite{ItkinKitapbayev2025}.

By applying the value-matching condition $P(t,x) = K - X_B(t)$ at $x = X_B(t)$ to equation \eqref{decompDensity} with $H(u, X_u)$ given by \eqref{Htd}, we can derive a nonlinear Volterra integral equation of the second kind for the optimal exercise boundary $X_B(t)$ (note that this approach generalizes to the broader case covered by \eqref{decompGen}), yielding
\begin{align} \label{Volterra}
K - X_B(t) &= e^{-r(T-t)} \int_0^K \left(K - X_T\right) \psi\left(X_T, T | X_B(t), t \right) d X_T \\
&+ \int_t^T e^{-r(u-t)} \Big[ r(u) K \Psi_1(u, X_B(u) | t, X_B(t)) - q(u) \Psi_2(u, X_B(u) |t,X_B(t)) \Big]  \mathbf{1}_{H(u, X_u) > 0}  du, \nonumber \\
\Psi_1(u, X_B(u)  | t,x) &=  \int_0^{X_B(u)} \psi\left(X_u, u | x, t \right) d X_u, \qquad
\Psi_2(u, X_B(u)  | t,x) =  \int_0^{X_B(u)} X_u \psi\left(X_u, u | x, t \right) d X_u. \nonumber
\end{align}
Therefore, an explicit knowledge of the transition density allows solving \eqref{Volterra} numerically, thus obtaining $X_B(t)$. Then the American Put option price can be found by calculating the integrals in the right-hand side of \eqref{decompDensity}.

For the American Call option $C(t,x)$ a representation similar to \eqref{decompDensity} can be derived by using the Call-Put symmetry, \cite{CarrJarrowMyneni1992}, which reads
\begin{align} \label{decompDensityCall}
C \left(t, x \right) &= D(t,T) \int_K^\infty \left(X_T - K\right) \psi\left(X_T, T | x, t \right) d X_T \\
&+ \int_t^T D(t,u) \int_{X_B(u)}^\infty H(u, X_u) \psi\left(X_u, u | x, t \right) d X_u d u, \qquad
H(u, X_u) \equiv q X_u - r K. \nonumber
\end{align}
The condition $H(u, X_u) < 0$ should be supplemented by $X_u > K$. Due to this Call-Put symmetry, this paper focuses on Put options, as analogous results for Call options can be derived using the symmetry.

\paragraph{Computing Greeks.} It should be noted that the notation $P(t, x)$ is somewhat loose, as it does not explicitly reflect the option price’s dependence on the model parameters, even though such dependence exists. Having this implicitly in mind to avoid ambiguity, we differentiate both sides of \eqref{decompDensity} with respect to a model parameter $y$, obtaining
\begin{align} \label{greekGen}
Y_A(t,x,y) &= Y_E(t,x,y) + \int_t^T D(t,u) \Bigg[  \int_0^{X_B(u)} H(u, X_u) \psi_y \left(X_u, u | x, t \right)  d X_u \\
&+ H(u, X_B(u)) \psi \left(X_B(u), u | x, t \right) \fp{X_B(u)}{y} \Bigg] du. \nonumber
\end{align}
Here $Y_A(t,x,y) = P_y\left(t, x \right)$ is the first order Greek of the American Put option price, and
\begin{equation}
Y_E(t,x,y) = D(t,T)\int_0^K \left(K - X_T\right) \psi_y\left(X_T, T | x, t \right) d X_T
\end{equation}
\noindent is the corresponding European Greek. Thus, computing the integral on the right-hand side of \eqref{greekGen} requires knowledge of $(X_B(u))'_y$. Notably, when $y = x$ (i.e., when computing the option Delta $\Delta$), we have $(X_B(u))'_y = 0$, simplifying the expression. The same applies to the option Gamma $\Gamma$, reducing \eqref{greekGen} to
\begin{align} \label{gamma}
\Gamma_A(t,x) &= \Gamma_E(t,x) + \int_t^T D(t,u) \int_0^{X_B(u)} H(u, X_u) \psi_{xx} \left(X_u, u | x, t \right)  d X_u du
\end{align}
\noindent with $\Delta_A(t, x \in \mathcal{E}) = -1$ and $\Gamma_A(t, x \in \mathcal{E}) = 0$. Also, since $X_B(t)$ is known from solving \eqref{Volterra}, further $X'_B(t)$ can be easily computed and substituted into \eqref{greekGen} to provide the option Theta $\Theta$.

For other Greeks, such as the option Vega, the derivative $(X_B(u))'_y$ is not known explicitly but can be determined by solving another integral equation. This equation can be derived by differentiating both sides of \eqref{Volterra} with respect to $y$ yielding the following Fredholm integral equation for $(X_B(t,u))'_y$
\begin{align} \label{greekEB}
Y_E(t,X_B(t,y)) &+ \int_t^T D(t,u) \int_0^{X_B(u,y)} H(u, X_u) \psi_y \left(X_u, u | X_B(t,y), t \right) dX_u du \\
&= - \int_t^T D(t,u) H(u, X_B(u,y)) \psi \left(X_B(u,y), u | X_B(t,y), t \right) \fp{X_B(u,y)}{y} du, \nonumber
\end{align}
\noindent subject to the terminal condition $(X_B(T,u))'_y = 0$. By selecting an appropriate numerical method (such as suitable quadrature techniques), this approach becomes more computationally efficient than computing Greeks through bumping model parameters and revaluing option prices, as has been commonly done in prior literature.

\subsection{An alternative approach}

Unfortunately, explicit solutions for the transition density are rare, especially for time-inhomogeneous models. Therefore, we can apply the alternative approach developed in \cite{ItkinMuravey2024jd}. Specifically, the authors derive Volterra integral equations of the second kind for various popular financial models in time-inhomogeneous settings, which can be solved to determine the exercise boundary (EB), the European option price and the option price in the continuation region.

\paragraph{Computing the EB.}
The methodology considers a PDE for the call option price in the continuation region
\begin{equation} \label{PDE}
\fp{C}{t} + \dfrac{1}{2}\Sigma^2(t, x) \sop{C}{x} +  \mu(t,x) \fp{C}{x} = r(t) C,
\end{equation}
\noindent subject to the terminal condition at option maturity $t=T$
\begin{equation} \label{tc0}
C(T,x) = (x-K)^+,
\end{equation}
\noindent and the boundary conditions:
\begin{equation} \label{bc0}
C(t,0) = 0, \qquad \Delta(t,X_B(t)) = 1, \qquad \Delta(t,x) = \fp{C(t,x)}{x}.
\end{equation}
As demonstrated in \cite{ItkinMuravey2024jd}, the terminal condition \eqref{tc0} in the continuation region becomes homogeneous: $C(T,x) = 0$.

Under this framework, the American Call option price $C(t,x)$ in the region $x \in [0, X_B(t)]$ coincides with the price of an Up-and-Out barrier option having the same maturity $T$, strike $K$, and time-dependent upper barrier $H = X_B(t)$. When the upper barrier is reached, the option terminates and the holder receives a rebate equal to $X_B(t) - K$.

This barrier option pricing problem admits a semi-analytical solution, as demonstrated in \cite{ItkinMuravey2024jd,ItkinLiptonMuraveyBook}. The resulting explicit representation of the Call option price (the same approach applies to Put options) depends on integrals that are functions of the yet-unknown EB and model parameters. By differentiating this representation with respect to $x$ and applying the smooth-pasting condition from \eqref{bc0}, we obtain a nonlinear integral equation for the EB of the type
\begin{equation} \label{nonLinVoltOU}
\xi(\tau, y(\tau)) = \int_0^\tau \mathcal{K}(\tau, s, y(s)) ds,
\end{equation}
\noindent where $\xi(\tau, y(\tau))$ is a known function of its arguments, $\tau$ represents the transformed time $t$, $y(\tau)$ is the transformed EB $X_B(t)$, and $\mathcal{K}(\tau, s, y(s))$ is an explicitly known kernel function specific to each model. For comprehensive details, see \cite{ItkinMuravey2024jd}.

Numerical solution of \eqref{nonLinVoltOU} yields the exercise boundary for the problem. Thus, when the transition density is not explicitly available, we can employ the methodology of \cite{ItkinMuravey2024jd} instead of using \eqref{Volterra} to determine the exercise boundary - the first essential component of the American option pricing framework.

\paragraph{The European option price.}
The second component involves computing the price of the corresponding European option. This follows a similar approach to the barrier option case, but with modified boundary conditions: the continuation region becomes $x \in [0, l_x)$, the terminal condition remains as in \eqref{tc0}, and the boundary condition at $x = l_x$ is specified as $C(t,l_x) = l_x - K$.

This European option pricing problem can be solved using the same methodology employed for the barrier option. The solution is obtained analytically by combining the generalized integral transform (GIT) technique with Duhamel's principle, which provides an explicit representation of the Green's function for the pricing PDE within the continuation region. For detailed treatment of this approach, again see \cite{ItkinMuravey2024jd} and also the final section of \cite{ItkinKitapbayev2025}.

\paragraph{Computing the EEP.}
The last piece of the puzzle is calculation of the EEP given the EB $X_B(t)$. Suppose, that we price the Put option and that the drift is given by \eqref{Htd} (this assumption, of course, can be easily relaxed). Also, we assume that $X_B(t)$ can be determined independently by using the approach of \cite{ItkinMuravey2024jd}, and thus, it is treated as a known function of time.

Looking again at \eqref{decompDensity}, one can observe that \eqref{decompDensity} in this particular case can be represented in the form
\begin{align} \label{decompDensityHedge}
P \left(t, x, T \right) &= P_E \left(t, x, T \right) + \int_t^T D(t,u) \left[ \bar{r}_u P_{\CON}(u, X_B(u)) - \bar{q}
P_{\AON}(u, X_B(u)) \right] du,
\end{align}
\noindent where $P_{\CON}(u, X_B(u)), P_{\AON}(u, X_B(u))$ are Cash-or-nothing Put option and Asset-or-nothing Put options, respectively, with maturity $u$ and strike $K = X_B(u)$. Alternatively, $P_{\CON}(u, X_B(u))$ can be treated as a barrier option with a {\it digital} payoff and the upper barrier being $X_B(t)$. Then $P_{\AON}(u, X_B(u))$ is a similar barrier {\it Call} option with zero strike.

Aside from hedging, if one needs to numerically compute the integral in \eqref{decompDensityHedge}, a more efficient approach is to treat each integrand as the price of a double barrier Call option with:
\begin{itemize}
\item Upper barrier: $X_B(u)$,
\item Maturity: $u$,
\item Payoff: $H(u, X(u))$,
\item Strike: zero,
\item rebate at $x=0$ equal to $r(t)K$.
\end{itemize}
By making a change of variables
\begin{equation}
\bar{C}(t,x) = C(t,x) - r(t)K,
\end{equation}
\noindent we reduce this problem to pricing an Up-and-Out barrier option with the following specifications: upper barrier $X_B(t) - r(t)K$, maturity $u$, strike $K$, payoff $H(u, X(u)) - r(t)K$, and rebate-at-hit equal to $H(u, X_B(u)) - r(t)K$ when the upper barrier is reached. Pricing formulas for such barrier options under various time-dependent models have been developed in \cite{ItkinLiptonMuraveyBook} and related papers of the authors, thus allowing us to reuse these results by substituting the appropriate parameters.

Since the integral in \eqref{decompDensityHedge} can be approximated by a finite sum over time $u$, we need a series of barrier option prices corresponding to each time point $u$. However, since these options differ only in their EB $X_B(u)$ and maturity $u$, those prices can be computed sequentially in time. This approach leverages prior calculations of the temporal integral up to $u - \Delta u$, requiring only new computations for the final time step $\Delta u$. Consequently, the computational procedure scales linearly with respect to $u$. In some cases, this may require expanding the Green's function into a series in $X_B(t)$ and using, for example, two terms to achieve $O((\Delta t)^2)$ accuracy, which represents standard tolerance for modern pricing models,  see, e.g., an example in \cref{s:numEx}.

Although we only mention this approach as a possibility in this particular case, in \cite{Nunes2020a} a similar methodology was adapted to the valuation of American-style double knock-out options with justification of its efficiency. Additionally, these authors investigated the pricing of American-style options under the JDCEV model of \cite{CarrLinetsky2006}, which considers diffusion with killing; see \cite{Nunes2009,Ruas2013,Nunes2020b}. We do not address this extension here.

\section{From transition density to characteristic function} \label{s:CF}

It is well-known that for many popular diffusion models, such as the Heston model, \cite{Rouah2013}, the transition density is not available in closed form, whereas the CF is. The canonical approach for pricing American options using the CF is the CONV algorithm, as described in \cite{Lord2008Siam}. This algorithm discusses two methods for valuing American options: a) approximating the American option as a Bermudan option with a large number of exercise opportunities, and b) applying Richardson extrapolation to a series of Bermudan options with increasing exercise dates. While \cite{Lord2008Siam} primarily employs the Bermudan approximation, this method introduces errors due to the discrete exercise schedule. Furthermore, the Bermudan option price is computed iteratively over time with step size $\Delta t$. At each time step $t_m$, the option value is given by
\begin{equation}
V(t_m, x) = \max (E(t_m, x), C(t_m, x)),
\end{equation}
\noindent where $C(t_m,x)$ is the continuation value (a European option price) and $E(t_m,x)$ is the exercise value. While this representation suffices for pricing, it suffers from poor accuracy when computing option Greeks, especially near the EB. To mitigate this issue, a penalty method is recommended, \cite{Zvan1998,Halluin2004}. Overall, in this approach, the CF is used solely to compute the European option price $C(t_m, x)$ via inverse fast Fourier transform (FFT). This relies on the observation that the standard risk-neutral valuation formula for European options can be expressed as a convolution.

We now describe an alternative approach that combines the option price representation from Proposition~\ref{prop1} with the transition density expression via the CF derived in \cite{FangOosterlee2008}.

The core insight of \cite{FangOosterlee2008} stems from the observation that while the conditional density $f(y\mid x)$ is typically unknown for many financial models, the characteristic function is often available. Assuming that $f(y\mid x)$ has full support on the interval $[a, b] \in \mathbb{R}$ with $b > a$, we can replace it with its cosine expansion in $y$
\begin{equation}
f(y \mid x) = \sum_{k=0}^{\infty}{\vphantom{\sum}}' A_k(x) \cos \left(k \pi \frac{y-a}{b-a}\right),
\end{equation}
\noindent where coefficients of the expansion $A_k(x)$ are given by
\begin{equation}
A_k(x):=\frac{2}{b-a} \int_a^b f(y \mid x) \cos \left(k \pi \frac{y-a}{b-a}\right) d y,
\end{equation}
\noindent and $\sum{\vphantom{\sum}}'$ means that the first term in the summation is weighted by one-half.

It then follows that over the interval $[a,b]$, the coefficients $A_k(x)$ can be approximated by coefficients $F_k(x)$ such that $A_k(x) \approx F_k(x)$. The latter can be expressed through the CF $\phi(x)$ (the Fourier pair of the density)
\begin{equation}
F_k(x) \equiv \frac{2}{b-a} \Ree \left\{\phi\left(\frac{k \pi}{b-a}; x\right) \exp \left(-\iu \pi \frac{k a}{b-a}\right)\right\},
\end{equation}
\noindent where $\iu$ is the imaginary unit, $\Ree(\cdot)$ denotes the real part, and
\begin{equation}
\phi(\omega) = \int_\mathbb{R} e^{i x \omega} f(x) d x.
\end{equation}

Substituting this representation into \eqref{decompDensity} yields the European option price for the first term in \eqref{decompDensity} expressed via the standard COS method, \cite{FangOosterlee2008}, and the EEP for the second term which reads
\begin{align} \label{eepCF}
\mathrm{EEP}(t,x) &= \frac{2}{b-a} \sum_{k=0}^{\infty}{\vphantom{\sum}}' \int_t^T du D(t,u) \Ree \left\{\phi\left(\frac{k \pi}{b-a}; x\right) \exp \left(-\iu \pi \frac{k a}{b-a}\right)\right\} \\
&\cdot \int_0^{X_B(u)} H(u, X_u)  \cos \left(k \pi \frac{X_u-a}{b-a}\right) d X_u. \nonumber
\end{align}

Given a particular model for the underlying asset and the corresponding expression for $H(u,X_u)$, the final integral in \eqref{eepCF} can be computed either numerically or, in many cases, analytically. For instance, with linear drift and $H(u,X_u)$ given in \eqref{Htd}, this yields
\begin{align}
\int_0^{X_B(u)} & H(u, X_u)  \cos \left(k \pi \frac{X_u-a}{b-a}\right) d X_u =
\begin{cases}
F(u,k,\Delta), & k > 0 \\
K r(u) X_B(u) - \frac{1}{2} q(u) X_B^2(u), & k = 0.,
\end{cases}
\\
F(u,k,\Delta) &= \frac{\Delta}{\pi^2 k^2} \Bigg\{
\Delta q(u) \cos \left(k \pi \frac{a}{\Delta}\right) + \cos \left(k \pi\frac{a - X_B(u)}{\Delta}\right) \nonumber  \\
&+ k\pi \left[ K r(u) \sin \left(k\pi\frac{a}{\Delta}\right) - (X_B(u) q(u) + K r(u)) \sin \left(k \pi\frac{a - X_B(u)}{\Delta}\right) \right] \Bigg\}, \qquad \Delta = b - a \nonumber
\end{align}
with $\Delta = b - a$.

The formula in \eqref{eepCF} extends naturally to time-inhomogeneous models. For such models, we modify our notation to accommodate the time dependence:
\begin{equation}
\phi(\omega,t; x) = \int_\mathbb{R} e^{i y \omega} f(y, t; x) , dy,
\end{equation}
\noindent where the density $f(y, t; x)$ now explicitly depends on time $t$, reflecting the time-varying nature of the model parameters. While the characteristic function may not be available in closed form for such models, this limitation can be addressed since the temporal integral in \eqref{eepCF} requires numerical computation regardless. For example, \cite{CarrItkinMuravey2020} consider barrier option pricing for the time-inhomogeneous Heston model, which requires knowledge of the corresponding CF. While this expression is well-established in the time-homogeneous case, when model coefficients are time-dependent functions $p(t)$, the authors propose using the method of \cite{Guterding2018}.

The approach leverages the fact that for affine models - such as the Heston model — the CF admits a closed-form solution, provided the associated time-inhomogeneous Riccati equation can be solved. Therefore, the method involves partitioning the entire time interval $t \in [0,T]$ into $N$ subintervals of length $\Delta t = T/N$, then approximating time-dependent model parameters with piecewise constant coefficients. On each interval $i$, we have $p(t) = p_i$ for $i=1,\ldots,N$, where $p(t)$ represents a time-dependent model parameter and $p_i$ is its constant value on the $i$-th interval. The solution $\alpha_{i}(t,p)$ of the Riccati equation for each interval $i$ is obtained using the known solution for constant parameters, with the solution from the previous interval $\alpha_{i-1}(t,p)$ serving as the terminal condition. The time-dependent Riccati equation is solved backward in time, starting with the boundary condition $\alpha(T,p) = 0$. As demonstrated in \cite{Guterding2018,CarrItkinMuravey2020}, this procedure is fully analytical and very fast computationally.

Since the early exercise boundary $X_B(u)$ can be determined in advance using the approach described above, the temporal integral on the right-hand side of \eqref{eepCF} can be computed numerically using modern quadrature methods, such as double-exponential quadrature, etc. Furthermore, the COS method provides exponential convergence as the number of terms in the truncated sum $\sum{\vphantom{\sum}}'$ increases, making the overall approach both accurate and computationally efficient.

In should be noted that the COS method, however, could be unstable for small maturities and in some other cases. In this case, usage of the Non-Uniform Fast Fourier Transform (NUFFT) method could be preferable, see \cite{AndersenNUFFT} and discussion therein. Also, the described approach of dealing with integral equations for pricing American options can be further generalized for the case when, say the underlying stock pays discrete dividends, \cite{AndersenLake2025}.

\section{Multidimensional diffusion models} \label{s:nDim}

An additional advantage of the advocated approach is that the American option price representation similar to that in \eqref{decompDensity} applies to any diffusion model without jumps, i.e. models with multiple stochastic drivers or multidimensional models. This generality extends the method's applicability to pricing American options under stochastic volatility models and their various extensions.

Suppose that we consider a model
\begin{equation} \label{modelN}
dX^i_t = \mu_i(t, \bm{X}_t) dt + \Sigma_i(t, \bm{X}_t) dW^i_t, \qquad \bm{X} = \left(X^1, \ldots, X^n\right), \ i \in [1,n],
\end{equation}
\noindent and all Brownian motions $W^i_t$ are pairwise correlated with the constant correlation coefficient $\rho_{i,j} dt  = <d W^i_t, dW^j_t>$. Then, generalization of \eqref{decompDensity} can be rigorously stated by proposition, which generalizes Proposition~1 in \cite{ItkinKitapbayev2025}.  The proof relies on an extension of Peskir's change of variables formula, presented in \cite{Peskir2007Surf} and restated here for the reader's convenience.

\begin{proposition} \label{propS}

Let $\bm{X} = \left(X^1, \ldots, X^n\right)$ be a continuous semimartingale with $n \ge 1$ and let $b: \mathbb{R}^{n-1} \rightarrow \mathbb{R}$ be a continuous function such that the process $b^X=b\left(X^1, \ldots, X^{n-1}\right)$ is a semimartingale. Setting $C=\left\{\left(x_1, \ldots, x_n\right) \in \mathbb{R}^n \mid x_n<b\left(x_1, \ldots, x_{n-1}\right)\right\}$ and $D=\left\{\left(x_1, \ldots, x_n\right) \in \mathbb{R}^n \mid x_n>\right.$ $\left.b\left(x_1, \ldots, x_{n-1}\right)\right\}$ suppose that a continuous function $F: \mathbb{R}^n \rightarrow \mathbb{R}$ is given such that $F$ is $C^{i_1, \ldots, i_n}$ on $\bar{C}$ and $F$ is $C^{i_1, \ldots, i_n}$ on $\bar{D}$ where each $i_k$ equals 1 or 2 depending on whether $X^k$ is of bounded variation or not. Then the following change-of-variable formula holds
\begin{align}
F\left(\bm{X}_t\right) &= F\left(\bm{X}_0\right)+\sum_{i=1}^n \int_0^t \frac{1}{2}\left(\frac{\partial F}{\partial x_i}\left(X_s^1, \ldots, X_s^n+\right)+\frac{\partial F}{\partial x_i}\left(X_s^1, \ldots, X_s^n-\right)\right) d X_s^i \\
&+ \frac{1}{2} \sum_{i, j=1}^n \int_0^t \frac{1}{2}\left(\frac{\partial^2 F}{\partial x_i \partial x_j}\left(X_s^1, \ldots, X_s^n+\right)+\frac{\partial^2 F}{\partial x_i \partial x_j}\left(X_s^1, \ldots, X_s^n-\right)\right) d\left\langle X^i, X^j\right\rangle_s \nonumber \\
&+ \frac{1}{2} \int_0^t\left(\frac{\partial F}{\partial x_n}\left(X_s^1, \ldots, X_s^n+\right)-\frac{\partial F}{\partial x_n}\left(X_s^1, \ldots, X_s^n-\right)\right) I\left(X_s^n=b_s^X\right) d \ell_s^b(X), \nonumber
\end{align}
\noindent where $\ell_s^b(X)$ is the local time of $X$ on the surface $b$ given by
\begin{equation}
\ell_s^b(X)=\mathbb{P}-\lim _{\varepsilon \downarrow 0} \frac{1}{2 \varepsilon} \int_0^s  I\left(-\varepsilon<X_\tau^n-b_\tau^X<\varepsilon\right) d\left\langle X^n-b^X, X^n-b^X\right\rangle_\tau,
\end{equation}
\noindent and $d \ell_s^b(X)$ refers to integration with respect to $s \mapsto \ell_s^b(X)$. The analogous formula extends to general semimartingales $X$ and $b^X$ as well.

\begin{proof}[{\bf Proof}]
See \cite{Peskir2007Surf}.
\end{proof}
\end{proposition}

A key distinction between the multidimensional case (with multiple stochastic drivers) and the one-dimensional setting is that the exercise boundary now becomes an $(n-1)$-dimensional surface. For example, in a simple stochastic volatility model with two stochastic variables $(S_t, v_t)$, the boundary takes the form $b = X_B(t, v)$.

Also, when the diffusions are correlated, the quadratic variation terms are replaced by their covariation counterparts. Combined with the other bounded variation terms in the formula, they yield the infinitesimal generator of the model.

Based on the result of Peskir, the following proposition holds:
\begin{proposition} \label{prop3}
Under the assumptions of Proposition~\ref{propS}, conditional on $X^i_t = x^i, \ i=1,\ldots,n$, the American Put price with a {\bf single} exercise boundary $X_B(t,x_1,\ldots,x_{n-1})$ can be represented by the following decomposition formula
\begin{align} \label{decompGenN}
P \left(t, \bm{x} \right) &= \EQ \left\{ D(t,T) [K - X^n_T]^+ | {\bm X}_t = {\bm x}\right\} + \int_t^T D(t,u) \EQ \left\{ \left[ r(u)(K-X^n_u) + \mu_n(u,\bm{X}_u) \right] \mathbf{1}_{\bm{X}_u \in \mathcal{E}} \right\} d u,
\end{align}
\noindent where the spot price is marked as $X^n_t$.

\begin{proof}[{\bf Proof}]
Let us choose $F(t, \bm{X}_t) = D(t,T) P(T,\bm{X}_T)$. Then under some mild regularity conditions using \Ito lemma and change-of-variable formula in Proposition~\ref{propS}, the following representation holds
\begin{align} \label{changeVar}
D(t,T) &P(T,\bm{X}_T) = P(t,\bm{x}) + \int_t^T D(t,u) \Big[ \mathbb{L}_X P(u,\bm{X}_u) - r(u) P(u, \bm{X}_u) \Big] du + M_T + \frac{1}{2} \int_t^T D(t,u) \\
&\times \Big[ P_{x^n}(u, b(u)+) - P_{x^n}(u,b(u)-) \Big] d\ell_u^b(X), \qquad
M_u = \sum_i \int_t^u D(t,u) P_{x^i}(u, \bm{X}_u) \Sigma(u,\bm{X}_u) dW^i_u, \nonumber \\
\mathbb{L}_X f &:= f_t + \sum_i \mu_i(t,\bm{x}) f_{x^i} + \frac{1}{2} \sum_{i,j} \rho_{i,j} \Sigma_i(t,\bm{x}) \Sigma_j(t,\bm{x}) f_{x^i x^j}. \nonumber
\end{align}
Here, $\mathbb{L}_X f$ is an infinitesimal generator of \eqref{modelN}, $M_u$ is a martingale part of the transformation with $M_t = 0$. It is important that \eqref{changeVar} holds at the entire domain $(t,\bm{x}) \in \mathbb{R}_{+} \times \mathbb{R}^n \rightarrow \mathbb{R}_{+}$.

Due to the smooth-pasting condition for American options $P_{x^n}(t, b(t)\pm) = -1$, the integral over the local time in \eqref{changeVar} vanishes. Taking the expectation $\EQ$ of the remaining parts yields
\begin{align} \label{changeVar2}
D(t,T) &\EQ \left[ P(T, \bm{X}_T)\right] = P(t,\bm{x}) + \int_t^T D(t,u) \EQ \Big\{ \left[ \mathbb{L}_X P(u,\bm{X}_u) - r(u) P(u, \bm{X}_u)\right] \mathbf{1}_{\bm{X}_u \in \mathcal{C}} \Big\} du \\
&+ \int_t^T D(t,u) \EQ \Big\{ \left[ \mathbb{L}_X P(u,\bm{X}_u) - r(u) P(u, \bm{X}_u)\right] \mathbf{1}_{\bm{X}_u \in \mathcal{E}} \Big\}. \nonumber
\end{align}
The term in the left-hands part is the price of the corresponding European option. The first integral in the right-hand side of this equation vanishes due to the Feynman-Kac theorem valid in the continuation region $\bm{X}_u \in \mathcal{C}$. For the second integral, in the exercise region $P(u, \bm{X}_u) = K - X^n_u$, hence $\mathbb{L}_X P(u,X_u) = - \mu_n(u, \bm{X}_u)$. Finally, by rearranging the terms, we obtain \eqref{decompGenN}.
\end{proof}
\end{proposition}

To underline, the expectation $\EQ$ is now multidimensional. Therefore, for instance, in the case of a stochastic volatility model with two stochastic drivers, computation of $\EQ$ requires two integrations, one over the spot price and the other one over $v$.

Although the general decomposition formula in \eqref{decompGenN} appears novel, similar decomposition approach has been employed in several studies to price American options under the Heston stochastic volatility model with constant coefficients, which represents a particular case of our proposed framework, see e.g. \cite{Chiarella2005,AitSahalia2010} and references therein. It is interesting that, similar to our approach, in \cite{Chiarella2005} the authors express the transition density function of the corresponding PDE through the inverse Fourier transform of the characteristic function, leveraging the closed-form availability of this function for the Heston model. However, this approach seems to be less computationally efficient than our method, which utilizes the COS method. That is because the former requires FFT and has complexity $O(N \log N)$ where $N$ is the number of the FFT nodes, while the latter has an exponential convergence in $N$.

Since the EB for the Heston model is not known in closed form, \cite{Chiarella2005} determined it by assuming that the optimal stopping surface can be well-approximated in a log-linear fashion near the long-term variance level, i.e. as a linear function of the long-term variance model parameter. The coefficients of this regression were then determined using a coarse implementation of the LSMC. A more sophisticated approach to computation of the EB, but of the same nature is used in \cite{AitSahalia2010}.

In contrast to these methods, and following the framework established in the previous section, the EB can be determined by setting $\bm{X}_u = b(u)$ and $P(u, \bm{x}) = K - b(u, x_{1,u},\ldots,x_{n-1,u}$ in equation \eqref{decompGenN}, then solving the resulting integral Volterra equation of the second kind numerically. This approach is demonstrated for the one-dimensional case in \cite{ItkinKitapbayev2025}.

Furthermore, in \cite{CarrItkinMuravey2020} the GIT method is developed for pricing barrier options in the time-dependent Heston model, also with time-dependent barriers. Our approach represents the option price in a semi-analytical form as a two-dimensional integral. This methodology can be naturally extended to derive the corresponding 2D integral Volterra equation of the second kind for the EB (as demonstrated in \cite{ItkinMuravey2024jd} for various one-dimensional models)  since the Green's function of the corresponding PDE was already established in \cite{CarrItkinMuravey2020}.

Interestingly, a similar integral equation approach for the Heston model  was explored in \cite{Chiarella2006}, again using a linear approximation of the EB. However, the authors note that their method suffers from slow convergence, most likely due to an ill-conditioned system for the regression coefficients. They hypothesize that this ill-conditioning may arise from the integral equations' insensitivity to small perturbations in the free boundary.

\section{Models with jumps} \label{s:Jump}

In this section we describe another extension of the proposed approach for the models with jumps.  Again, the foundation for this approach can be found in \cite{Peskir2007Surf}, Section~3 which  considers semimartingales first with jumps of the bounded variables, and then generalizes this description for general semimartingales.

If $X=\left(X^1, \ldots, X^n\right)$ is a general semimartingale, then each $X^i$ can be decomposed into
\begin{equation} \label{eq32}
X_t^i=X_0^i+X_t^{i, c}+X_t^{i, d},
\end{equation}
\noindent with $X^{i, c}=M^{i, c}+A^{i, c}$ and $X^{i, d}=M^{i, d}+A^{i, d}$ where $M^{i, c}$ is a continuous local martingale, $A^{i, c}$ is a continuous process of bounded variation, $M^{i, d}$ is a purely discontinuous local martingale, and $A^{i, d}$ is a pure jump process of bounded variation.

Then, the following extension of Proposition~\ref{propS} is proven in  \cite{Peskir2007Surf}

\begin{proposition} \label{propJ}

Let $X=\left(X^1, \ldots, X^n\right)$ be a semimartingale, let $b: \mathbb{R}^{n-1} \rightarrow \mathbb{R}$ be a continuous function such that the process $b^X=b\left(X^1, \ldots, X^{n-1}\right)$ is a semimartingale, and let $F: \mathbb{R}^n \rightarrow \mathbb{R}$ be a continuous function such that $F$ is $C^{i_1, \ldots, i_n}$ on $\bar{C}$ and $F$ is $C^{i_1, \ldots, i_n}$ on $\bar{D}$ where each $i_k$ equals 1 or 2 depending on whether $X^k$ is of bounded variation or not. Recall the Tanaka formula in the form
\begin{align} \label{loctimeJ}
\left|X_t^n-b_t^X\right| &= \left|X_0^n-b_0^X\right|+\int_0^t \operatorname{sign}\left(X_{s-}^n-b_{s-}^X\right) d\left(X_s^n-b_s^X\right)+\ell_t^b(X) \\
&+ \sum_{0<s \leq t} \left( \left|X_s^n-b_s^X\right| - \left|X_{s-}^n-b_{s-}^X \right| - \operatorname{sign}\left(X_{s-}^n-b_{s-}^X\right) \Delta\left(X^n-b^X\right)_s \right), \nonumber
\end{align}
\noindent where $\operatorname{sign}(0)=0$.  Define $\ell_s^b(X)$ - the local time of $X$ on the surface $b$ - given by means of \eqref{loctimeJ}. Then the following change-of-variable formula holds
\begin{align} \label{CVJ}
F\left(X_t\right) &= F\left(X_0\right)+\sum_{i=1}^n \int_0^t \frac{1}{2}\left(\frac{\partial F}{\partial x_i}\left(X_{s-}^1, \ldots, X_{s-}^n+\right)+\frac{\partial F}{\partial x_i}\left(X_{s-}^1, \ldots, X_{s-}^n-\right)\right) d X_s^i \\
&+ \frac{1}{2} \sum_{i, j=1}^n \int_0^t \frac{1}{2}\left(\frac{\partial^2 F}{\partial x_i \partial x_j}\left(X_{s-}^1, \ldots, X_{s-}^n+\right)+\frac{\partial^2 F}{\partial x_i \partial x_j}\left(X_{s-}^1, \ldots, X_{s-}^n-\right)\right) d\left[X^{i, c}, X^{j, c}\right]_s \nonumber \\
&+ \sum_{0<s \leq t}\left(F\left(X_s\right)-F\left(X_{s-}\right)-\sum_{i=1}^n \frac{1}{2}\left(\frac{\partial F}{\partial x_i}\left(X_{s-}^1, \ldots, X_{s-}^n+\right)+\frac{\partial F}{\partial x_i}\left(X_{s-}^1, \ldots, X_{s-}^n-\right)\right) \Delta X_s^i\right) \nonumber \\
&+ \frac{1}{2} \int_0^t\left(\frac{\partial F}{\partial x_n}\left(X_{s-}^1, \ldots, X_{s-}^n+\right)-\frac{\partial F}{\partial x_n}\left(X_{s-}^1, \ldots, X_{s-}^n-\right)\right) I\left(X_{s-}^n=b_{s-}^X, X_s^n=b_s^X\right) d \ell_s^b(X), \nonumber
\end{align}
\noindent where $d \ell_s^b(X)$ refers to integration with respect to the continuous increasing function $s \mapsto \ell_s^b(X)$.

\begin{proof}[{\bf Proof}]
See \cite{Peskir2007Surf}, Theorem~3.2
\end{proof}
\end{proposition}

\begin{remark}
In \cite{Wilson2019}, Peskir's framework for semimartingales with jumps is further analyzed and extended. A key insight is that the "smooth fit" condition along the boundary \( b \) — used in deriving \eqref{decompGenN} — is significantly weaker than requiring \( F \) to be \( C^{1,2} \) or even \( C^1 \) at \( b \). Drawing on \cite{Toit2009}, the author presents additional and alternative conditions that are more readily verifiable, while also relaxing some original assumptions in \cite{Peskir2007Surf}. These refinements accommodate both continuous and discontinuous boundary processes \( b \), as well as the existence of local times for \LY processes (including those with jumps of bounded variation and \( \alpha \)-stable \LY processes. etc.).
\end{remark}

It is known that models with jumps introduce market incompleteness, meaning that a unique equivalent martingale measure (EMM) does not exist in this setting. Instead, multiple EMMs exist, all satisfying the no-arbitrage condition but yielding different option prices. To resolve this ambiguity, a common approach is to price options under a specific pricing measure $\mathbb{Q}$, which is determined by calibrating the model to observed market option prices; see \cite{ContTankov,ItkinBook} and references therein.

Consider the jump-diffusion process $X_t = (X_t)_{0 \le t \le T}$ of the form
\begin{equation} \label{stockEvol}
dX_t = \mu(t,X_t) dt + \Sigma(t,X_t) dW_t + dL_t,
\end{equation}
\noindent which is the sum of a drift, Brownian motion and a jump part defined via the infinitesimal generator of the pricing semigroup
\begin{align} \label{PIDE-23}
\mathbb{L}_X f(t,x) &= f_t + \mu(t,x) f_x + \frac{1}{2} \Sigma^2(t,x) f_{x,x} + \int_{R\backslash\{0\}} \left[ f(t,x e^y) - f(t,x) - x (e^y-1)\fp{f(t,x)}{x} \right] \nu(dy).
\end{align}
\noindent where $\mu \in \mathbb{R}$, $\Sigma \in \mathbb{R}^+$, $W_t$ is a standard Brownian motion, and $\nu(dy)$ is the density of jumps (\LY density), which satisfies all the existence conditions of the \LY measure, see discussion in \cite{Levendorskii2008}.

The existence of a smooth pasting condition for models with jumps has been thoroughly discussed in \cite{Kyprianou2005} and references therein. For example, \cite{BL2002} established sufficient conditions for a class of \LY processes with exponential moments and stable-like characteristic exponents (RLPE processes), under which continuous or smooth pasting occurs. For this class of processes, they showed that smooth pasting holds if \(X\) has bounded variation with a nonpositive drift term, or if \(X\) has unbounded variation. Their method also provides identities for the value function and optimal exercise barrier for more general claim structures and for \LY processes with absolutely continuous resolvents.

In \cite{PeskirShiryaev2002}, it was further observed for spectrally negative processes that smooth pasting fails if and only if the process is of bounded variation. A related work \cite{Chan2004} noted that in the case of the American put option, continuity still holds even when smooth pasting does not. Thus, \cite{PeskirShiryaev2002} offers an intuitive explanation for when and why smooth pasting fails in their models, advocating instead for a principle of continuous pasting.

In what follows, we assume that the jump models satisfy a smooth pasting condition, restricting our analysis to those models meeting the criteria outlined in \cite{Kyprianou2005}. Building on \eqref{CVJ}, we can prove another proposition analogous to Proposition~\ref{prop3}.
\begin{proposition} \label{propJmy}
Under a jump-diffusion model of type \eqref{stockEvol}, conditional on $X_t = x$, the American Put price with a single exercise boundary $X_B(t)$ solves the following integral Volterra equation of the second kind
\begin{align} \label{decompGenJ}
P \left(t, x \right) &= \EQ \left\{ D(t,T) [K - X_T]^+ | X_t = x\right\} + \int_t^T D(t,u) \EQ \left[ \Omega(u.X_u) \mathbf{1}_{X_u \in \mathcal{E}} \right] du, \\
\Omega(u.X_u) &= \mu(u,X_u) + r(u) (K - X_u) + \int^\infty_{\log(X_B(u)/X_u)} \left[ P_E(u, X_u e^y) - K + X_u e^y \right]  \nu(dy). \nonumber
\end{align}
Here, the first term represents the usual European Put price $P_E\left(t, x \right)$ while the second term is the EEP which depends on the early exercise boundary $X_B(t)$ and the \LY measure of the process $\nu(dy)$.

\begin{proof}[{\bf Proof}]
The proof follows the same steps as in Proposition~\ref{prop1} (or Proposition~\ref{prop3}), with the key difference that  the infinitesimal generator of the model now includes an additional jump term - the last one on the right-hand side of \eqref{PIDE-23}. Accordingly, under same regularity conditions, we choose $F(t, X_t) = D(t,T) P(T,X_T)$ and use \Ito lemma and change-of-variable formula, \cite{Peskir2007}, to get
\begin{align} \label{changeVar3}
D(t,T) &P(T,X_T) = P(t,x) + \int_t^T D(t,u) \Big[ \mathbb{L}_X P(u,X_u) - r(u) P(u, X_u) \Big] du + M_T + \frac{1}{2} \int_t^T D(t,u) \\
&\times \Big[ P_x(u, X_B(u)+) - P_x(u,X_B(u)-) \Big] d\ell_u(X;X_B). \nonumber
\end{align}
Due to the smooth-pasting condition for American options $P_x(t, X^*(t)\pm) = -1$, the integral over the local time in \eqref{changeVar3} vanishes. Taking the expectation $\EQ$ of the remaining parts yields
\begin{align} \label{changeVar2J}
D(t,T) &\EQ \left[ P(T,X_T)\right] = P(t,x) + \int_t^T D(t,u) \EQ \Big\{ \left[ \mathbb{L}_X P(u,X_u) - r(u) P(u, X_u)\right] \mathbf{1}_{X_u \in \mathcal{C}} \Big\} du \\
&+ \int_t^T D(t,u) \EQ \Big\{ \left[ \mathbb{L}_X P(u,X_u) - r(u) P(u, X_u)\right] \mathbf{1}_{X_u \in \mathcal{E}} \Big\}. \nonumber
\end{align}
The first integral in the right-hand side of this equation vanishes due to the Feynman-Kac theorem valid in the continuation region $X_u \in \mathcal{C}$. For the second integral, in the exercise region $P(u, X_u) = K - X_u$. In the second integral the term under the expectation side can be found by using the definition of $\mathbb{L}_X$ in \eqref{PIDE-23}. This yields
\begin{align} \label{genEx}
\Omega(u.X_u) &= - \Big[ \mathbb{L}_X P(u,X_u) - r(u) P(u, X_u) \Big] \mathbf{1}_{X_u \in \mathcal{E}} \\
&= \mu(u,X_u) + r(u) (K - X_u) + \int_{R\backslash\{0\}} \left[ P(u, X_u e^y)\mathbf{1}_{X_u \in \mathcal{E}} - K + X_u e^y \right]  \nu(dy). \nonumber
\end{align}
If $X_u e^y \in \mathcal{E}$, it implies that $P(u, X_u e^y) = K - X_u e^y$, and thus the integral in \eqref{genEx} vanishes. Therefore, the limits on integration in $y$ can be adjusted to follow $X_u e^y \in \mathcal{C}$. For the Put option that means that $X_u e^y > X_B(u)$. Therefore, the integration should be done for $y > \log (X_B(u)/X_u)$. Since in \eqref{genEx} $X_u \in \mathcal{E}$, this means that $X_B(u)/X_u \ge 1$ and $y > 0$. Thus, we have
\begin{equation} \label{omega}
\Omega(u.X_u) = \mu(u,X_u) + r(u) (K - X_u) + \int^\infty_{\log(X_B(u)/X_u)} \left[ P(u, X_u e^y) - K + X_u e^y \right]  \nu(dy).
\end{equation}
The inequality $X_u e^y > X_B(u)$ implies $P(u, X_u e^y) = P_E(u, X_u e^y)$. Substituting this expression into \eqref{changeVar2J}, we obtain \eqref{decompGenJ}.
\end{proof}
\end{proposition}

The EEP term in the form of \eqref{decompGenJ} has a clear economic interpretation, as noted by \cite{Chiarella2009}. The component $\mu(u,X_u) + r(u) (K - X_u)$ reflects the net cash flows from holding the portfolio $K-X_u$ when $X_u \in \mathcal{E}$, i.e. a concept familiar from the pure diffusion case. The integral term, however, arises solely due to jumps in the $X_t$ process. This additional term accounts for the rebalancing costs faced by the option holder when the underlying price jumps upward from the stopping region back into the continuation region.

For instance, suppose the underlying price $X_u$ is at $X_u(-) < X_B(u)$ during the option's life, placing it in the stopping region. The option holder thus holds the portfolio $K-X_u$. If a jump of size $y$ occurs at $\tau > 0$ such that $X_u(+) = y X_u(-) > X_B(\tau)$, the portfolio's value falls above that of the unexercised American Put. The integral in \eqref{omega} precisely captures this cost differential.

Again, while the general expression \eqref{decompGenN} for processes with jumps appears to be novel, certain special cases, particularly jump models with constant coefficients, have been studied in the literature. Specifically, the decomposition approach has been applied to pricing American options in jump-diffusion settings. For instance, \cite{Lamberton2013}
studies the behavior of the EB of an American Put option near maturity in an exponential \LY model. In particular, it proves that in situations where the limit of the EB is equal to the strike price, the rate of convergence to the limit is linear if and only if the underlying \LY process has finite variation. In the case of infinite variation, a variety of rates of convergence can be observed: we prove that when the negative part of the \LY measure exhibits an $\alpha$-stable density near the origin, with $1<\alpha<2$, the convergence rate is ruled by $\theta^{1 / \alpha}|\ln \theta|^{1-\frac{1}{\alpha}}$, where $\theta$ is the time until maturity.

To emphasize an alternative approach, \cite{Itkin2024jd} developed a method for computing the EB in time-dependent jump-diffusion models with exponential and double exponential jumps. Their framework: i) accommodates arbitrary time-dependent model parameters; ii) reduces the American option pricing problem to solving a nonlinear algebraic equation for the EB and a linear Fredholm-Volterra integral equation for the option's Gamma; iii) derives option Greeks by solving analogous Fredholm-Volterra equations obtained via differentiation of the pricing equation. As noted earlier, solving integral equations rather than partial integro-differential equations (PIDEs) typically yields superior accuracy at comparable speeds or faster computation for equivalent accuracy.

In \cite{Chiarella2009}, the American option pricing problem is considered in the case where the underlying
asset follows a jump-diffusion process, namely the Merton model which describes jumps of finite activity and finite variation. They apply a version of Jamshidian transformation to convert a time-homogeneous PIDE valid in the continuation region to time-inhomogeneous PIDE valid in the whole space domain. This PIDE now includes a new integral time-inhomogeneous term which actually represents  the EEP. Then they solve it using same technique as in \cite{Chiarella2005}, i.e. by applying the Fourier transform to this PIDE, using Duhamel's principle to address a source term (via a series representation) and computing expectations using the inverse Fourier transform and the CF of the process instead of the transition density. As we have already mentioned, this approach seems to be computationally involved.

\subsection{Determining the EB} \label{s:EB}

Determining the EB requires knowing the EB value just prior to expiry, which is essential for numerically solving the derived equations. While the literature often assumes this limit coincides with that of the pure diffusion case, \cite{Chiarella2009} shows that the reality is more nuanced. By analyzing the inhomogeneous term in the PIDE, the authors demonstrate that jumps significantly influence the EB at expiry. Their work provides an explicit analytical derivation of this difference. A more comprehensive study of this problem across various jump-diffusion models can be found in \cite{Levendorskii2008}, along with related works by the same authors cited therein.

This analysis can be reproduced in our case of a general jump-diffusion model as well. Indeed, in \cite{Chiarella2009} the derivation of the limit of $x_B(t)$ as $t \to T^-$ is done based on their another paper \cite{Chiarella2004} which extends the analysis of \cite{Wilmott2000}. They show that the latter is equivalent to setting the inhomogeneous term of their PDE (which is the EEP) to zero, setting $t=T, X=X_B(T^-)$, and solving for the EB. In our case this is equivalent to solving the equation $Q(u,X_u) = 0$.

\begin{proposition} \label{EBatTJ}
Denote $X_B(T^-) = b$. Then the limit of the EB $X_B(t)$ in the jump-diffusion model with the infinitesimal generator $\mathbb{L}_X$ given in \eqref{PIDE-23}, at $t=T^-$ is given by
\begin{align} \label{fPropEBatT}
b &= K \min (1, b_*),
\end{align}
\noindent where $b_*$ solves the equation
\begin{align} \label{EBatT}
\frac{b_*}{K} &= \frac{ r(T) - \int^\infty_{\log(K/b_*)} \nu(dy) }{r(T) - \int^\infty_{\log(K/b_*)} e^y \nu(dy)} + \frac{\mu(T,b_*)/K}{r(T) - \int^\infty_{\log(K/b_*)} e^y \nu(dy)}.
\end{align}
In particular, when $\mu(t,X_t) = [ r(t) - q(t)]X_t$, \eqref{EBatT} reduces to
\begin{align} \label{EBatT1}
\frac{b_*}{K} &= \frac{ r(T) - \int^\infty_{\log(K/b_*)} \nu(dy) }{q(T) - \int^\infty_{\log(K/b_*)} e^y \nu(dy)}.
\end{align}

\begin{proof}[{\bf Proof}]
As mentioned above, the value of $b$ can be determined by solving the equation
\begin{equation} \label{omega0}
\mu(u,X_u) + r(u) (K - X_u) + \int^\infty_{\log(X_B(u)/X_u)} \left[ P(u, X_u e^y) - K + X_u e^y \right]  \nu(dy) = 0,
\end{equation}
\noindent evaluated at $u=T$ and $X_u = b$. Using the Put option payoff $P(T, x) = (K - x)^+$, this becomes
\begin{equation} \label{omega01}
\mu(T,b) + r(T) (K - b) + \int^\infty_{\log(K/b)} \left[ \max(K - b e^y,0) - K + b e^y \right]  \nu(dy) = 0,
\end{equation}
\noindent where we also took into account that $X_B(T) = K$. Since the lower integration limit in $y$ is $\log(K/b)$, the maximum function in \eqref{omega01} vanishes, further simplifying the expression to obtain
\begin{equation} \label{omega02}
\mu(T,b) + r(T) (K - b) - \int^\infty_{\log(K/b)} \left[K - b e^y \right]  \nu(dy) = 0.
\end{equation}

This can be rearranged to the form
\begin{equation} \label{omega03}
b  = K \Bigg[ \frac{ r(T) - \int^\infty_{\log(K/b)} \nu(dy) }{r(T) - \int^\infty_{\log(K/b)} e^y \nu(dy)} + \frac{\mu(T,b)/K}{r(T) - \int^\infty_{\log(K/b)} e^y \nu(dy)} \Bigg].
\end{equation}
For instance, if $\mu(t,b) = [r(t) - q(t)] b$, this yields
\begin{equation} \label{omega04}
b  = K \frac{ r(T) - \int^\infty_{\log(K/b)} \nu(dy) }{q(T) - \int^\infty_{\log(K/b)} e^y \nu(dy)},
\end{equation}
\noindent (compare with a particular case of this formula for the American Call option in \cite{Chiarella2009J}). Finally, since for the American Put option we have $X_B(t) \le K$, combining this condition with \eqref{omega04} yields \eqref{fPropEBatT}.
\end{proof}
\end{proposition}

\subsection{Method of pseudo-differential operator}

To determine $X_B(t)$, one must numerically solve a Volterra integral equation of the second kind which follows from Proposition~\ref{propJ}, by setting $X_t = X_B(t)$ in \eqref{decompGenJ}
\begin{equation} \label{VolEBJ}
K - X_B(t) = P_E\left(t, X_B(t)\right) + \int_t^T D(t,u) \mathbb{E} \left[ \Omega(u,X_u) | X_t = X_B(t) \right] du.
\end{equation}
This equation is challenging to solve since, by the definition of $\Omega(u,X_u)$ in \eqref{omega}, the integral on the right-hand side of \eqref{VolEBJ} depends on $P(u,X_u e^y)$, making it \emph{non-local} in $P(u,X_u)$. Even when $X_B(t)$ is known (found), solving the integral equation in \eqref{decompGenJ} for the Put price would encounter similar difficulties.

One possible resolution is to apply the method of \cite{ItkinBook}, which converts this non-local operator to a \emph{local} operator of an unusual nature — specifically, a \emph{pseudo-differential} operator. To demonstrate this, let us
consider the jump part of $\mathbb{L}_X P(u,X_u)$ in \eqref{changeVar2J} which, based on \eqref{PIDE-23}, reads
\begin{equation}
J = \int_{\mathbb{R}} \left[ P(u,X_u e^y) - P(u,X_u) - X_u (e^y-1)\fp{P(u,X_u)}{X_u} \right] \nu(dy).
\end{equation}

By changing variables in this integral $X_u \mapsto z_u = \log X_u$ and $y \mapsto \bar{y} = \log(y)$ we obtain
\begin{equation} \label{jumpI}
J = \int_{\mathbb{R}} \left[ P(u,z_u + \bary) - P(u,z_u) - (e^y-1)\fp{P(u,z_u)}{z_u} \right] \nu(d\bary).
\end{equation}

It is well known from quantum mechanics \cite{OMQM} that a translation (shift) operator in $L_2$ space can be represented as
\begin{equation} \label{transform}
\mathcal{T}_b = \exp \left( b \frac{\partial}{\partial x} \right),
\end{equation}
\noindent with $b$ constant, so that
\begin{equation} \label{shift}
\mathcal{T}_b f(x) = f(x+b).
\end{equation}

This can be easily proved using a Taylor series expansion of both sides of \eqref{shift}. Therefore, the integral in \eqref{jumpI} can be represented as (see \cite{ItkinBook} for a more detailed discussion and examples)
\begin{align} \label{jumpI2}
J &= \opJ P(u,z_u), \qquad \opJ \equiv \int_{\mathbb{R}} \left[ e^{\bary \nabla} + 1 - (e^y-1)\nabla \right] \nu(d\bary), \qquad  \nabla = \fp{}{z_u}.
\end{align}

Given a \LY measure $\nu(d\bary)$, the integral $\opJ$, can be formally computed under mild assumptions about existence and convergence if one treats $\nabla$ as a constant. Therefore, operator $\opJ$ can be considered as a generalized function of the differential operator $\partial_{z_u}$ or as a pseudo-differential operator.

As shown in \cite{ItkinBook}, for many popular jump models, the \LY density is such that the operator $\opJ$ can be found in closed form. Moreover, the following proposition holds
\begin{proposition}[Proposition~5.6 from \cite{ItkinBook}] \label{cfJump}
Suppose we are given a \LY process $X_t$ with $\psi(\omega)$ being the characteristic exponent of $z_t = \log X_t$. Then
\begin{equation*}
\opJ = \psi(-\iu \nabla).
\end{equation*}

\begin{proof}
 This directly follows from the \LY- Khinchin theorem, \cite{ContTankov}. \hspace*{\fill} $\square$
\end{proof}
\end{proposition}
Accordingly, an alternative representation for $\Omega(u.X_u)$ can be proposed in the form
\begin{align} \label{omegaNew}
\Omega(u.X_u) &= \mu(u,X_u) + r(u) (K - X_u) + \psi(-\iu \nabla) [P(u, z_u(X_u))\mathbf{1}_{X_u \in \mathcal{E}} ] \\
&=  \mu(u,X_u) + [r(u) + \psi(-\iu \nabla)] \left(K - e^{z_u(X_u)} \right). \nonumber
\end{align}
Then, since this pseudo-differential operator is {\it local}, it can be simplified by locally approximating it using various \pade approximations of the necessary tolerance, see an example in Appendix~\ref{app1}. Thus, this approach reduces \eqref{VolEBJ} to a simpler nonlinear equation with respect to $X_B(t)$.

\section{Numerical examples}  \label{s:numEx}

This section provides numerical illustrations of both the GIT and decomposition methods, demonstrating their application to computing  the EB through several representative examples.

\subsection{The GIT method}

To illustrate the GIT method, we follow the approach in \cite{ItkinMuravey2024jd} and consider a time-dependent Ornstein-Uhlenbeck (OU) model where the underlying spot price $S_t$ follows the stochastic differential equation (SDE):
\begin{equation} \label{OU1}
dS_t = [r(t) - q(t)] \, S_t dt + \sigma(t)dW_t.
\end{equation}

The corresponding integral equation for determining the EB was derived in \cite{ItkinMuravey2024jd}. Following their specification, we parameterize the time-dependent OU model as:
\begin{equation} \label{ex}
r(t) = r_0, \qquad q(t) = q_0 e^{-q_k t}, \qquad \sigma(t) = \sigma_0 e^{-\sigma_k t},
\end{equation}
\noindent where $r_0, q_0, q_k, \sigma_0, \sigma_k$ are constants. This parametrization enables an analytical solution to the corresponding Riccati equation presented in \cite{ItkinMuravey2024jd}. The model parameters used in our numerical experiments are shown in Table~\ref{tab1}\footnote{In these tests we use $\sigma_k > 0$, so the volatility is decreasing in time. However, there is no restriction on this value, so negative values of $\sigma_k$ can be used as well.}.
\begin{table}[!htb]
\begin{center}
\begin{tabular}{|c|c|c|c|c|c|}
\hline
$r_0$ & $q_0$ & $\sigma_0$ & $q_k$ & $\sigma_k$ & $T$  \\
\hline
0.02 & 0.01 & 2$K$ & 0.01 & 10 & 1 \\
\hline
\end{tabular}
\caption{Model parameters for numerical tests.}
\label{tab1}
\end{center}
\end{table}
Bear in mind that $\sigma(t)$ represents the normal volatility. We obtain its value by multiplying the log-normal volatility by the strike level $K$, which is a standard practice for normal models.
\begin{figure}[!htp]
\begin{center}
\subfloat[]{\includegraphics[width=0.54\textwidth]{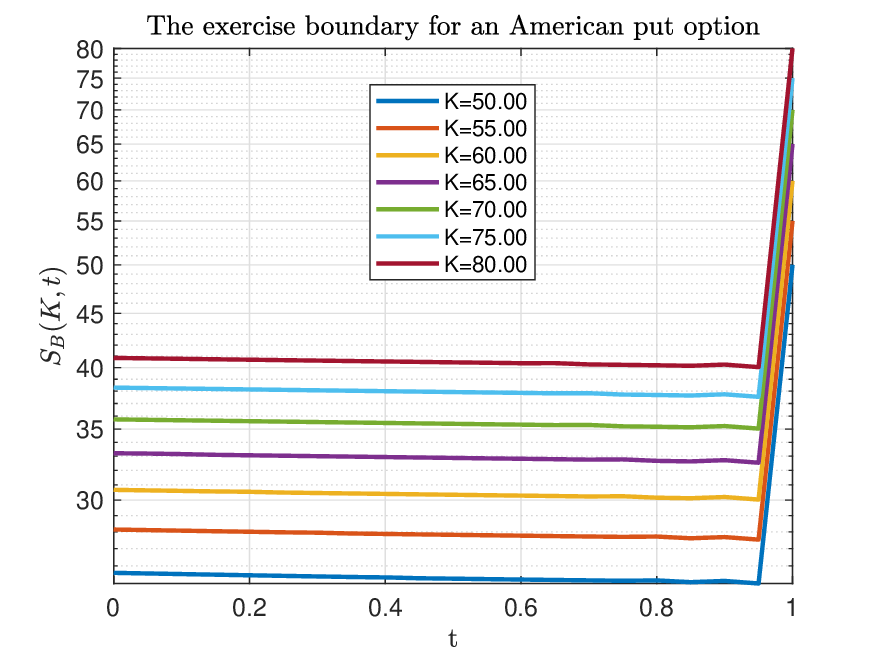}}
\hspace*{-0.3in}
\subfloat[]{\includegraphics[width=0.54\textwidth]{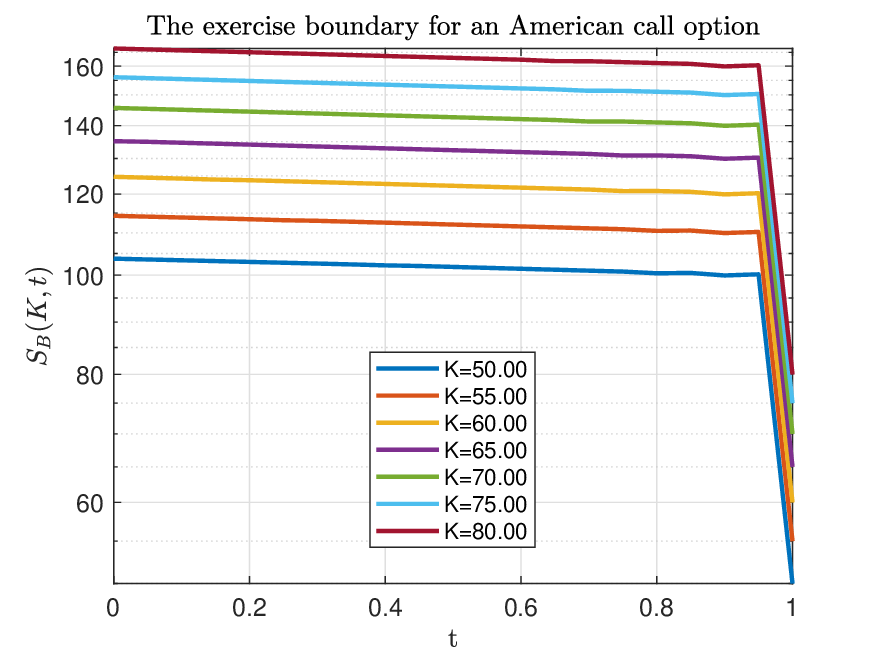}}
\end{center}
\caption{Semi-log plot of the exercise boundary $S_B(t)$ for American put and call options under the time-dependent OU model for various strikes $K \in [50,55,60,65,70,75,80]$: (a) Put options with model parameters from Table~\ref{tab1}; (b) Call options with identical parameters except $r_k = 0.01, q_k = 0$.}
\label{OUtest}
\end{figure}

We conduct numerical tests for a range of strikes $K \in \{50, 55, 60, 65, 70, 75$, $80\}$. Fig.~\ref{OUtest}(a) presents the EB for American Put options as functions of time $t$. The boundaries are computed by solving the corresponding Volterra equation from \cite{ItkinMuravey2024jd} using trapezoidal quadrature. Higher-order quadrature methods such as Simpson’s rule, which achieves \( O(\Delta t)^4) \) accuracy in time, or adaptive integration schemes can further enhance the precision of our approach while requiring only minimal implementation changes and no algorithmic modifications. In contrast, achieving similar accuracy improvements with FD or LSMC methods remains challenging.

We run the test for a set of strikes $K \in [50, 55, 60, 65, 70, 75, 80]$. The previously computed value $y(t_{i-1})$ is used as the initial guess, and the iterative method typically converges within 3-4 iterations. The computational efficiency is noteworthy: computing $S_B(t)$ on a temporal grid $t_i = i\Delta t$, $i \in \{0, 1, \ldots, M\}$, $\Delta t = T/M$ for all strikes with $M = 20$ requires only 0.33 seconds in MATLAB on a system with two Intel Quad-Core i7-4790 CPUs at 3.80 GHz. While this MATLAB implementation could be naturally vectorized for improved performance, we have not pursued this optimization since our focus is on demonstrating the basic methodology, which can be enhanced in various ways as discussed in \cite{ItkinMuravey2024jd}.

Fig.~\ref{OUtest}(b) shows analogous results for American call options using the same model parameters as in Table~\ref{tab1}, except with $r_k = 0.01$ and $q_k = 0$. It is worth noting that due to the put-call symmetry relation, \cite{Kwok2022}, for this model we have
\begin{equation}
S^{\text{call}}_B(t; r, q)\, S^{\text{put}}_B(t; q, r) = K^2.
\end{equation}

\subsection{The decomposition method}

We consider the 3/2 model introduced in \cite{Heston_32,Lewis:2000} which is a non-affine stochastic volatility model whose analytical tractability has been extensively studied, see \cite{CarrSun,ItkinCarr2010,Drimus2012} among others. Under this framework, the stock price $S_t$ follows the dynamics
\begin{align} \label{32model}
\frac{\mathrm{d} S_t}{S_t} &= [r(t) - q(t)] dt+\sqrt{V_t}\left(\rho \mathrm{~d} W_t^1+\sqrt{1-\rho^2} \mathrm{~d} W_t^2\right) \\
\mathrm{d} V_t &= V_t\left(\theta(t)-\kappa V_t\right) \mathrm{d} t+\varepsilon V_t^{3 / 2} \mathrm{~d} W_t^1, \nonumber
\end{align}
\noindent where $W_t^1$ and $W_t^2$ are independent Brownian motions under the risk-neutral measure $\mathbb{Q}$.

The 3/2 model exhibits a different behavior from the Heston model's square-root process with affine drift. Most notably, the mean reversion speed $\kappa V_t$ is linear in $V_t$, making it a stochastic quantity rather than a constant parameter. This structure implies that when $\kappa > 0$ (the typical case), mean reversion accelerates during periods of high instantaneous variance, creating a natural stabilizing mechanism. The volatility of variance parameter $\varepsilon$ also requires careful interpretation. Unlike its Heston counterpart, $\varepsilon$ in the 3/2 model must be scaled by $V_t$ to achieve comparable vol-of-variance effects. This scaling reflects the model's non-affine structure and contributes to its distinctive volatility clustering properties.

In the standard formulation, the long-term mean reversion level equals $\theta/\kappa$ for constant $\theta$. However, as pointed out by \cite{ItkinCarr2010}, $\theta_t$ could be an independent stochastic process in the most general setting. By conditioning on the path of $\theta_t$, the analytic tractability of the $3/2$ model with stochastic mean reversion level remains intact. Although in practice $\theta_t$ is often taken to be constant, this flexibility allows for more delicate modelling of the instantaneous variance dynamics when necessary. Here, following \cite{ZhengZheng2016}, we adopt the intermediate approach of time-dependent deterministic $\theta(t)$, which balances analytical tractability with enhanced modeling capability for time-varying market conditions.

As with the Heston model, parameter restrictions are necessary to ensure model well-posedness. Specifically, the conditions that guarantee both the non-explosion of the variance process $V_t$ and the martingale property of the discounted asset price. According to \cite{Drimus2012}, the 3/2 model parameters must satisfy the constraint
\begin{equation}
\kappa - \rho \varepsilon \ge - \frac{1}{2} \varepsilon^2
\end{equation}
This condition is typically satisfied under standard market conditions where $\kappa > 0$ (positive mean reversion) and $\rho < 0$ (negative correlation between asset returns and volatility, reflecting the leverage effect).

The joint CF of the log-price $x_u = \log S_u$ and $v_u$ for the 3/2 model is known in closed form, \cite{Lewis:2000,CarrSun,ZhengZheng2016}, and reads
\begin{align} \label{CF32}
\phi(u, &\omega, \lambda \mid t, x, v) \equiv \EQ[e^{\iu \omega x_u + \iu \lambda v_u}] \\
&= e^{\iu \omega x + \iu \omega a(t,u) } v^{-\alpha} \frac{\Gamma(\gamma - \alpha)}{\Gamma(\gamma)} \left( \frac{2}{\varepsilon^2 z(t,u)} \right)^\alpha M\left( \alpha, \gamma, -\frac{2}{\varepsilon^2 z(t,u) v} \right), \nonumber
\end{align}
where
\begin{align}
\alpha &= \frac12 - \frac{\kappa}{\varepsilon^2} + \sqrt{ \left( \frac12 - \frac{\kappa}{\varepsilon^2} \right)^2 + 2 \frac{\iu a(t,u) \rho + \lambda}{\varepsilon^2} + \frac{a^2(t,u)}{\varepsilon^2} }, \\
\gamma &= 2\left( \alpha + \frac{\kappa}{\varepsilon^2}\right),
\quad z(t,u) = \int_t^u e^{\int_t^s \theta(\tau) d\tau} ds, \quad a(t,t') = \int_t^{t'} (r(s) - q(s) ds. \nonumber
\end{align}
Here $\Gamma$ denotes the Gamma function, and $M(a,b,z)$ represents the Kummer confluent hypergeometric function, \cite{as64}.

From Proposition~\ref{prop3}, it follows that the EB in this model satisfies the integral Volterra  equation of the second kind
\begin{align} \label{Vol32}
K &- P_E(t,x_B(t,v),v) = e^{x_B(t,v)} + \int_t^T du\, D(t,u) \int_0^\infty dv_u \\
&\cdot \int_{-\infty}^{x_B(u, v_u)} dx_u\, \left[ r(u) K - q(u) e^{x_u} \right] \psi(u,x_u, v_u | t, x_B(t,v), v), \nonumber
\end{align}
\noindent where $x_u = \log X_u, \, x_B(t,v) = \log X_B(t,v)$, and $\psi$ is the joint transition density of $x_u, v_u$. Using the double COS expansion, this density can be represented as
\begin{align} \label{psi2}
\psi(t, x_u, &v_u \mid x, v, t) \approx \frac{4}{L_x L_v} \sum_{k=0}^{N_x-1}{}^{'} \;    \sum_{m=0}^{N_v-1}{}^{'} \Ree\Bigl[ e^{-\iu \bigl( \frac{k\pi a_x}{L_x} + \frac{m\pi a_v}{L_v} \bigr)} \\
&\cdot \phi\Bigl( \frac{k\pi}{L_x},\; \frac{m\pi}{L_v} \mid x, v, t \Bigr) \Bigr]
\cos\Bigl( \frac{k\pi (x_u-a_x)}{L_x} \Bigr) \cos\Bigl( \frac{m\pi (v_u-a_v)}{L_v} \Bigr), \nonumber
\end{align}
\noindent with $\phi\Bigl( \frac{k\pi}{L_x},\; \frac{m\pi}{L_v} \mid x, v, t \Bigr)$ given in \eqref{CF32}.

Moreover, the integration over $x_u$ in \eqref{Vol32} can be done in closed form. For doing that, let us substitute \eqref{psi2} into \eqref{Vol32} and re-write it in the form
\begin{align} \label{Vol32_1}
K &- P_E(t, x_B(t,v),v) = e^{x_B(t,v)} + \frac{4}{L_x L_v} \sum_{k=0}^{N_x-1}{}^{'} \; \sum_{m=0}^{N_v-1}{}^{'} \int_t^T du\, D(t,u) \int_0^\infty dv_u \nonumber \\
&\cdot \Ree\left[ e^{-\iu\left( \frac{k\pi a_x}{L_x} + \frac{m\pi a_v}{L_v} \right)} \phi\left( \frac{k\pi}{L_x},\; \frac{m\pi}{L_v} \mid x,v,t \right) \right] J(k,m),
\end{align}
where
\begin{align}
J(k,m) &= \int_0^\infty \cos\left( \frac{m\pi (v_u-a_v)}{L_v} \right) g_k(v_u), \\
g_k(v_u) &= \int_0^{x^*(u,v_u)} \left( r_d K - r_f e^{x_u} \right) \cos\left( \frac{k\pi (x_u-a_x)}{L_x} \right) d x_u \nonumber \\
&= r(u) K F_k(x^*(v)) - q(u) \Ree\left[ e^{-\iu\frac{k\pi a_x}{L_x}} \cdot \frac{e^{\alpha_k x^*(v)} - 1}{\alpha_k} \right]. \nonumber
\end{align}
Here, $\alpha_k = 1 + \iu\frac{k\pi}{L_x}$, and
\begin{equation}
F_k(x^*) =
\begin{cases}
x^*, & k = 0, \\
\frac{L_x}{k\pi} \left[ \sin\left( \frac{k\pi (x^*-a_x)}{L_x} \right) - \sin\left( -\frac{k\pi a_x}{L_x} \right) \right], & k > 0.
\end{cases}
\end{equation}

The series in \eqref{Vol32_1} are truncated to retain only the first $N_x$ and $N_v$ terms. As noted in \cite{FangOosterlee2008}, the cosine series expansion of entire functions (i.e., functions analytic everywhere in the complex plane except at infinity) converges exponentially. Consequently, for functions without singularities on \([a, b]\), the truncated series provides highly accurate approximations even for modest values of $N_x, N_v$.

The Volterra equation in \eqref{Vol32_1} can be solved backward in time subject to the terminal condition. Based on the analysis in \cref{s:EB},  it reads $X_B(T^-,v) = K \min[1, r(T)/q(T)]$. Moreover, since the CF of the model is known, $P_E(t,x,v)$ can also be efficiently computed using the standard COS method.

In a test example, the  constant parameters of the 3/2 model in \eqref{32model} are taken from \cite{Garshasebi2023}, who calibrated the 3/2 model with jumps to  short-term options on the SPX index. The time-dependent parameters $r(t), q(t) \theta(t)$ of the model, without loss of generality, are parameterized  as follows
\begin{equation} \label{ex3/2}
r(t) = r_0 + r_1 e^{- r_2 t}, \qquad q(t) = q_0 + q_1 e^{-q_k t}, \qquad \theta(t) = \theta_0 + \theta_1(T-t),
\end{equation}
\noindent where $T$ denotes the option's maturity. The specific parameter values are provided in Table~\ref{tab2}
\begin{table}[!htb]
\begin{center}
\begin{tabular}{|c|c|c|c|c|c|c|c|c|c|c|c|}
\hline
$\kappa$ & $\varepsilon$ & $\rho$ & $r_0$ & $r_1$ & $r_2$ & $q_0$ & $q_1$ & $q_2$ & $\theta_0$ & $\theta_1$ \\
\hline
22.84 & 8.56 & -0.9 & 0.005 & 0.02 & 2 & 0 & 0.015 & 0.5 & 0.467 & 0.05 \\
\hline
\end{tabular}
\caption{Parameters of the time-dependent 3/2 model used for numerical tests.}
\label{tab2}
\end{center}
\end{table}

\begin{figure}[!htp]
\begin{center}
\includegraphics[width=0.54\textwidth]{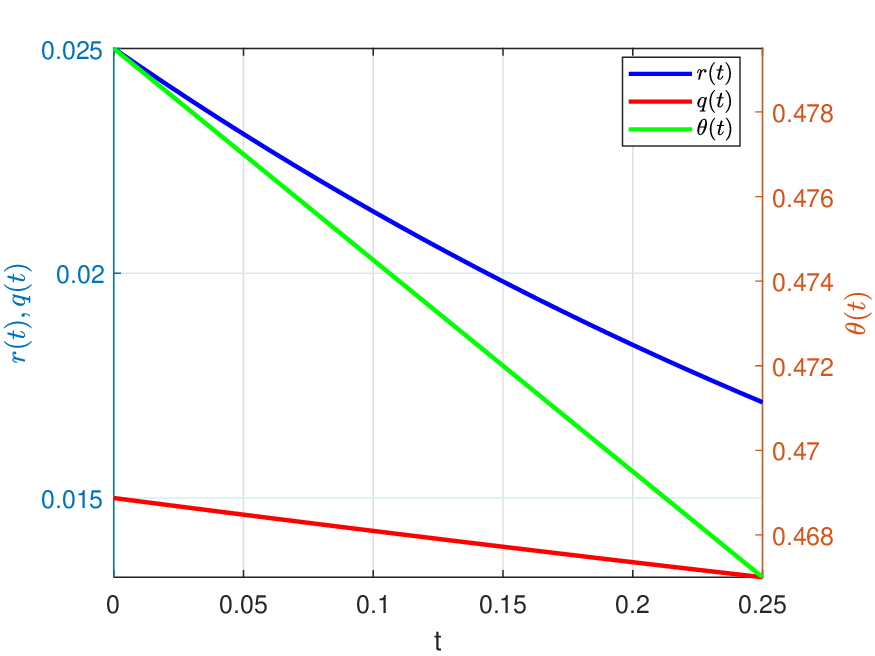}
\end{center}
\caption{The behaviour of time-dependent parameters of the 3/2 model for $ t \in [0,T]$.}
\label{rqTheta}
\end{figure}

For this test, we set the initial parameters to $S_0 = 100, K = 100, T = 0.25$. The dynamics of the model’s time-dependent parameters are illustrated in Fig.~\ref{rqTheta}.

We solve \eqref{Vol32_1} by discretizing the right-hand-side integral using trapezoidal quadrature with 26 evenly spaced points $t_i \in [0,T]$ (creating 25 intervals). Following \cite{FangOosterlee2008}, we set the domain truncation parameters as $b = -a = 10$ to adequately capture the characteristic function's cumulants. The solution proceeds backward in time from the terminal condition $X_B(T^-,v) = \min(1, r(T)/q(T))$. At each discrete time point $t_i$ ($i = 1,\dots,25$), we determine $X_B(t_i,v)$ by solving \eqref{Vol32_1}.

Computing $J(k,m)$ can be done very efficiently using a discrete cosine transform (DCT). The computational complexity of the method is: (a) computing the internal integral in $x$ (analytic evaluation): $O(N_x N_v)$; (b) computing DCT for each $k$: $O(N_x N_v \log N_v)$; (c) final summation: $O(N_x N_v)$; so, in total, $O(N_x N_v \log N_v)$. This is far better than the naive $O(N_x N_v \cdot N_{\text{quad}}^2)$, where $N_{\text{quad}}$ refers to the number of quadrature points used for numerical integration in each dimension if we were to compute the double integral using numerical quadrature (e.g., Gaussian quadrature or Simpson's rule).


We implement our algorithm in MATLAB, using the \verb|fsolve| function to solve the nonlinear equation for $X_B(t_i)$. For each step, we use $X_B(t_{i+1},v)$ as the initial guess, which typically leads to convergence within 3-5 iterations, achieving an absolute tolerance of approximately $10^{-12}$. For the COS method implementation, we consider two cases; with
$N_x = N_v = 32$ the elapsed time for the entire curve $X_B(t_i,v), t_i \in [0,T], \, i \in [1,25]$ is 25 seconds, while with $N_x = N_v = 64$ the elapsed time increases to 58 seconds.

The computational time is primarily driven by the evaluation of complex-valued special functions in \eqref{CF32}. For comparison, in the Heston model, where the CF includes two exponential terms, one logarithm, and one square root, the execution time decreases significantly to just 0.08 seconds.  For industrial applications, further speed improvements can be achieved by employing the NUFFT, since according to \cite{AndersenNUFFT}, the NUFFT is approximately 40 times faster than the COS method. Furthermore, implementing the entire algorithm in C++ reduces computation time to the microsecond range.

\begin{figure}[!htp]
\centering
\subfloat[3D representation]{\includegraphics[width=0.54\textwidth]{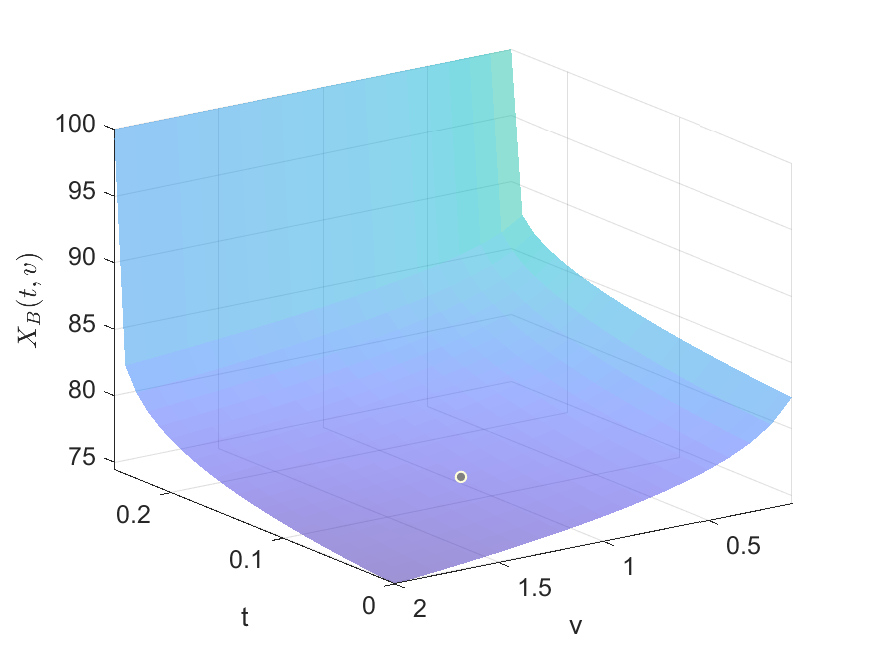}} \hspace*{-0.3in}
\subfloat[2D projections]{\includegraphics[width=0.54\textwidth]{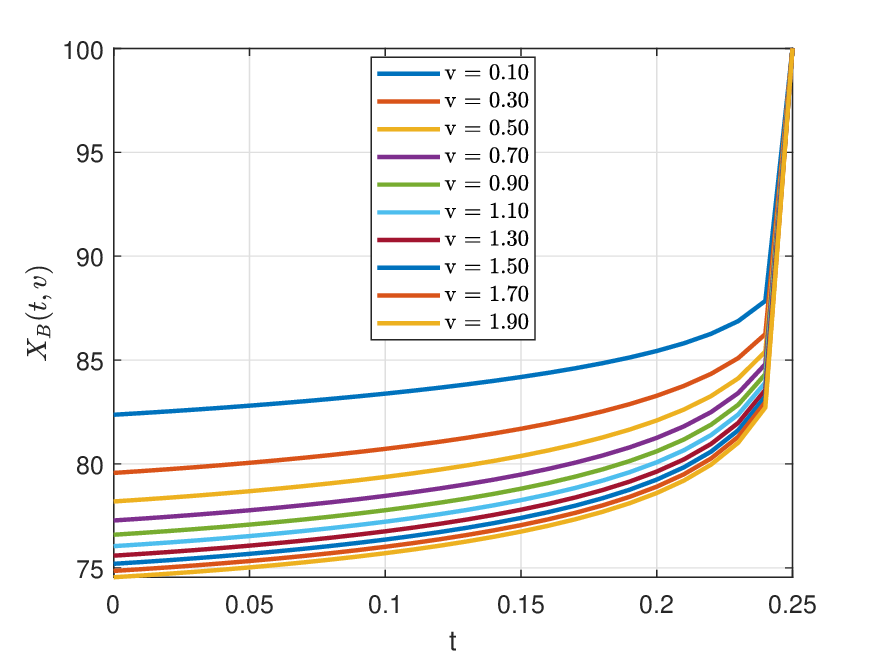}}
\caption{Early exercise boundary $X_B(t,v_t)$ for American put options under the time-dependent 3/2 model. Panel (a) shows the complete 3D surface, while panel (b) displays 2D cross-sections for various variance values $v_t$, computed using parameters from Table~\ref{tab2}.}
\label{res32}
\end{figure}

Since the EB in the 3/2 model is a function of both time \( t \) and variance \( v_t \), we evaluate $X_B(t, v)$ over the range \( v_t \in [0.1, 1.9] \). Fig.~\ref{res32} displays the resulting EB surface for different values of \( v_t \), computed using the parameters specified in Table~\ref{tab2}.

It can be seen that the shape of the EB over time is influenced by both, the temporal variation of the model parameters and the value of $v_t$. This dependence is particularly pronounced at low variance levels ($v_t \approx 0.1$), while the EB becomes nearly invariant to $v_t$ at higher variance values ($v_t \approx 1.9$).
\begin{figure}[!htp]
\centering
\subfloat[$P(K,v)$]{\includegraphics[width=0.54\textwidth]{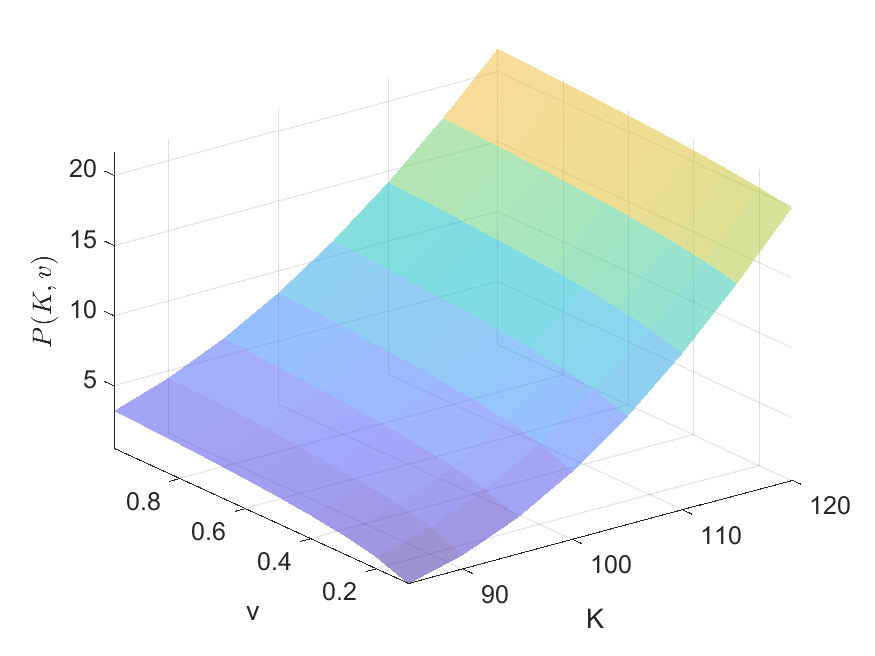}}
\hspace*{-0.3in}
\subfloat[$EEP(K,v)$]{\includegraphics[width=0.54\textwidth]{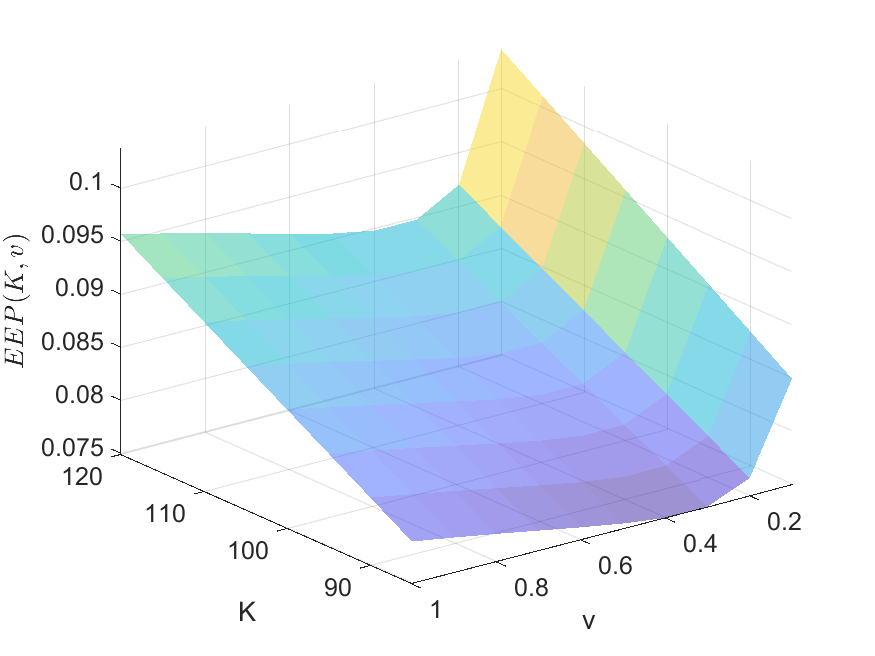}}
\caption{a) - American Put option price and (b) - the EEP,  under the time-dependent 3/2 model at various strikes $K$ and variances $v_t$, computed using parameters from Table~\ref{tab2}.}
\label{res32_1}
\end{figure}

Fig.~\ref{res32_1} displays the corresponding American Put option price along with the EEP. They are computed based on \eqref{eepCF} using the same test parameters and the EB calculated earlier and shown in Fig.~\ref{res32}. The EEP exhibits a parabolic dependence on $v_0$, reaching its minimum approximately at $v_0=0.35$ which lies below the mean-reversion level $\theta(t), \forall \in [0,T]$.

Note that a similar approach for the 3/2 model with constant coefficients has been explored in \cite{DetempleKitapbayev2017}, albeit using the transition density function. Additionally, an infinite-series representation for the transition density of the time-inhomogeneous 3/2 model was derived in \cite{Meesa2025}, where all coefficients are expressed in terms of Bell polynomials and generalized Laguerre polynomials.

\section{Conclusion}

In this paper, we use the change-of-variable formulas from \cite{Peskir2005,Peskir2007Surf} to derive general Volterra integral equations of the second kind for pricing American options under a broad class of {\it time-inhomogeneous} models. These models include pure diffusion processes (including multidimensional cases such as stochastic volatility models), jump-diffusion models, and models with jumps of finite or infinite variation and activity.

For models with jumps, the resulting integral equations are {\it non-local}, posing significant computational challenges. To address this, we demonstrate how these equations can be transformed into their {\it local} counterparts, which are more tractable, by using the pseudo-differential operator approach introduced in \cite{ItkinBook}.

Since for many such models the transition density is not known in closed form, but the CF is available, we propose representing the density via the CF using the representation provided by the COS method of \cite{FangOosterlee2008}. Once this representation is established, computing the option price requires two main steps: first, solving the integral Volterra equation (potentially multidimensional) of the second kind to find the EB, and then computing the European option price (usually via FFT) and the EEP using numerical quadratures.

Alternatively, the EB can be found by using the GIT approach described in \cite{ItkinMuravey2024jd,Itkin2024jd}, which also results in solving integral Volterra equation(s) of the second kind, but often with a more explicit (and therefore more easily computed) kernel.

Solving integral equations is typically faster and more accurate compared to other methods for pricing American options, such as FD, LSMC, etc. See the discussion in \cite{ItkinLiptonMuraveyBook,Andersen2016} and references therein for further details.

Regarding the novelty of our approach, we thoroughly analyzed the existing literature and found that a similar decomposition technique has previously been developed for only three {\it time-homogeneous} models: the Heston model in \cite{Chiarella2005}, the Merton jump-diffusion model in \cite{Chiarella2009}, and the 3/2 model in \cite{DetempleKitapbayev2017}. However, these studies do not employ the CF; instead, they either use the density directly (when available in closed form) or rely on the inverse FFT, which is computationally more demanding when standard methods are applied.

Our framework thus offers two key advancements: (i) a substantial generalization of the decomposition approach, and (ii) consequently, enhanced computational efficiency and accuracy. That said, we acknowledge that, as noted in \cite{AndersenNUFFT}, the COS method may exhibit instability for short maturities or certain other scenarios. In such cases, the NUFFT method proposed in \cite{AndersenNUFFT} could serve as a preferable alternative.

\section*{Acknowledgments}

I am grateful to Prof.~Goran Peskir and Leif Andersen for their valuable comments and discussions.

\printbibliography[title={References}]

\vspace{0.4in}
\appendixpage
\appendix
\numberwithin{equation}{section}
\setcounter{equation}{0}

\section{Using \pade approximation to compute a jump integral in \eqref{omegaNew}} \label{app1}

To illustrate this technique, we examine the Normal Inverse Gaussian (NIG) Model, introduced by \cite{NIG} as a model for stock price log returns. The NIG model is a subclass of the more general family of hyperbolic \LY processes. \cite{NIG} considered normal variance-mean mixtures and defined the NIG distribution as the case where the mixing distribution is inverse Gaussian, with the characteristic exponent
\begin{equation} \label{NIG_CE}
\Psi_{NIG}(\alpha, \beta, \delta, u) = \delta \left( \sqrt{\alpha^2 - \beta^2} - \sqrt{\alpha^2 - (\beta + \iu u)^2} \right).
\end{equation}
Thus, the CF is given by
\begin{equation} \label{NIG_CF}
\phi_{NIG}(\alpha, \beta, \delta, \mu, u) = \exp\left\{ t \delta \left( \sqrt{\alpha^2 - \beta^2} - \sqrt{\alpha^2 - (\beta + \iu u)^2} \right) + \iu t u \mu \right\},
\end{equation}
\noindent where $u \in \mathbb{R}, \ \mu \in \mathbb{R} \ \delta > 0, \ 0 \le |\beta| \le \alpha$.

The parameters of the NIG distribution have the following roles: $\alpha$ is responsible for the tail heaviness of steepness, $\beta$ affects the symmetry, $\delta$ scales the distribution, and $\mu$ determines the mean value (location). Also, it is  known that in option pricing applications, the location parameter \( \mu \) has no effect on the option value, so we may set \( \mu = 0 \) without loss of generality. However, this assumption is not critical for our approach and can be relaxed if needed.

The linearity of the log characteristic function in time indicates that the NIG process is infinitely divisible with stationary, independent increments. Since it doesn't have a diffusion component, the NIG process is a pure jump process with \LY density
\begin{equation} \label{NIGLevydens}
\nu(dx) = \dfrac{2 \alpha \delta}{\pi} \dfrac{\exp(\beta x) K_1(\alpha |x|)}{|x|} dx,
\end{equation}
\noindent where \( K_\lambda(z) \) is the modified Bessel function of the third kind (also known as the MacDonald function, \cite{BesselK}).

Using Proposition~\ref{cfJump} and \eqref{NIG_CF}, we obtain
\begin{equation} \label{JING}
\opJ = \delta \left( \sqrt{\alpha^2 - \beta^2} - \sqrt{\alpha^2 - (\beta + \triangledown)^2} \right).
\end{equation}
Thus, we need to compute
\begin{align}
a(u, X_u) &= \opJ(K - X_u) = \delta \left( \sqrt{\alpha^2 - \beta^2} - \sqrt{\alpha^2 - (\beta + \triangledown)^2} \right) \left( K - e^{z_u} \right),
\end{align}
\noindent where to recall, $z_u = \log X_u$. Here, parameters of the \LY density are assumed constant, though they could be generalized to, for example, piecewise constant functions of time \( u \).

Recall that $\beta, \alpha \in \mathbb{R}$. Observe that in our case, the operator $\opG := (\beta + \nabla)^2$ acts on the space of exponential functions, which also serve as its eigenfunctions. Specifically, for the operator $\opG$, $e^{z_u}$ is an eigenfunction with eigenvalue $(\beta + 1)^2$. If $\alpha^2 > (\beta + 1)^2$, the operator $\alpha^2 + \opG$ has a positive spectrum. Moreover, since $X_u \in \mathcal{E}$, it follows that $e^{z_u}$ is bounded. Consequently, the operator $\alpha^2 + \opG$ acting on a class of exponential functions is well-defined, meaning it admits a series expansion in $\nabla$. This holds despite the multi-valued nature of the square root function, and the expansion is expected to converge.

In \cite{ItkinBook}, the discretization of this operator on a FD grid in $z_u$-space is examined, along with the construction of a stable FD scheme for the corresponding PDE using a similar analysis of the properties of the above operator, but in the discrete space.

Once the well-defined behaviour of $\opJ$ is established, we can employ various approximations to compute $a(u,X_u)$. For instance, using a $(2,2)$ \pade rational approximation yields the equation
\begin{gather} \label{ODE}
a(u,z_u(X_u)) = \delta \frac{a_1 \nabla + a_2 \nabla^2}{b_1 \nabla + b_2 \nabla^2} \left(K - e^{z_u}\right), \\
\begin{alignat}{2}
a_1 &= 4 \sqrt{\alpha ^2 - \beta ^2} \left(\alpha ^4 \beta -3 \alpha^2 \beta^3 + 2 \beta^5 \right), \qquad&
a_2 &= 2 \sqrt{\alpha^2 - \beta^2} \left(\alpha^4 -3 \alpha^2 \beta^2 + 4 \beta^4 \right), \nonumber \\
b_1 &= - 2 \beta \left(\alpha^4 - 5 \alpha^2 \beta^2 + 4 \beta^4 \right), \qquad&
b_2 &= 4 \left(\alpha^2 - 2 \beta^2 \right) \left(\alpha^2 - \beta^2 \right)^2 - \alpha^4, \nonumber
\end{alignat}
\end{gather}
\noindent which, by setting $y(z_u) \equiv a(u,X_u(z_u))$,  can be re-written in the form
\begin{equation}
b_1 y'(z_u) + b_2 y''(z_u) = - \delta ( a_1 + a_2) e^{z_u}.
\end{equation}
This is an ordinary differential equation for $y(z_u)$ to be solved subject to the initial condition
\begin{equation}
y(\log K) = 0.
\end{equation}
This yields
\begin{equation}
a(u,X_u) = y(z_u) = \delta \frac{a_1 + a_2}{b_1 + b_2} (K - X_u) + C_1 \frac{b_2}{b_1}\left(K^{- b_1/b_2} - X_u^{- b_1/b_2}\right),
\end{equation}
\noindent where $C_1$ is the second constant of integration. If $b_1/b_2 < 0$, it can be found from the condition $y(z_u \to -\infty) = 0$, which directly follows from \eqref{ODE}. Solving this, we obtain
\begin{equation}
C_1 = - \delta \frac{a_1 + a_2}{b_1 + b_2} \frac{b_1}{b_2} K^{\frac{b_1 + b_2}{b_2}}.
\end{equation}
Otherwise, we need to set $C_1 = 0$.

\end{document}